\documentclass[showpacs,preprintnumbers,amsmath,amssymb,aps, prd]{revtex4}
\usepackage{graphicx}% Include figure files
\usepackage{grffile}
\usepackage{mathrsfs, mathtools}
\usepackage{caption}
\usepackage{float}
\usepackage{xcolor}
\usepackage{changepage}
\usepackage{dcolumn}% Align table columns on decimal point
\usepackage{bm}% bold math
\usepackage{graphicx,subfigure,epsfig}
\usepackage{multirow}

\def\mn{_{\mu\nu}}
\def\MN{^{\mu\nu}}

\def\b{\beta}
\def\f{\frac}

\def\c{\cite}

\def\rm{r_\mathrm{MBO}}
\def\ri{r_\mathrm{ISCO}}

\def\ei{E_\mathrm{ISCO}}
\def\lm{L_{MBO}}
\def\li{L_\mathrm{ISCO}}

\def\s{Schwarzschild }

\def\veff{V_{\text{eff}}(r)}
\newcommand\be{\begin{equation}}
\newcommand\ee{\end{equation}}
\newcommand\ba{\begin{eqnarray}}
\newcommand\ea{\end{eqnarray}}
\newcommand\bt{\bibitem}
\newcommand\nn{\nonumber}
\newcommand\lt{\left}
\newcommand\rt{\right}
\newcommand\pt{\partial}
\newcommand\tx{\text}
\newcommand\mc{\mathcal}
\begin{document}
\title{Equatorial periodic orbits and gravitational waveforms in Bardeen black holes surrounded by perfect fluid dark matter}
% Force line breaks with \
\author{Sohan Kumar Jha}
\email{sohan00slg@gmail.com}
\affiliation{Department of Physics, APC Roy Govt. College, Siliguri, West
Bengal, India}

\date{\today}% It is always \today, today, % but any date may be explicitly specified
\begin{abstract}
\begin{center}
Abstract
\end{center}
To probe the interplay between dark matter (DM) and non-linear electrodynamics (NED), we consider the Bardeen black hole (BH) surrounded by perfect fluid dark matter (PFDM). We first compute the effective potential governing the particle trajectory, and then, by imposing suitable conditions on the potential, examine the effects of DM and NED on the marginally bound orbit (MBO) and innermost stable circular orbit (ISCO). In this study, we confine the particle's trajectory to the equatorial plane. We then investigate periodic orbits around the Bardeen BH surrounded by PFDM (BPFDM BH), considering the rational number $q$ associated with each periodic orbit. We use the $(z,w,v)$ taxonomy, which is widely used to systematically organize periodic orbits. We examine the variation of $q$ with energy and angular momentum, and also the variation of the angular momentum and energy required for a specific $(z,w,v)$ configuration with the magnetic charge $g$ and DM parameter $\b$. Finally, with the help of the numerical "Kludge" method, we examine gravitational waveforms emitted from EMRIs where the central supermassive BH is modeled as a BPFDM BH. Our study reveals distinct signatures of NED and DM on orbital dynamics and gravitational waveforms.

\end{abstract}
\maketitle
\section{Introduction}
Despite remarkable accuracy of GR showcased by EHT's shadow observations of supermassive BHs (SMBHs) $M87^{*}$ and $SgrA^{*}$ \c{m87, sgra}, one issue that plagues the Einstein's general relativity (GR) is the existence of singularity which was shown inevitable when massive objects gravitationally collapse \c{sw1970}. Owing to singularity, geodesics remain incomplete, scalar invariants diverge, and as a consequence, physical laws lose their predictive power. This unavoidable feature of GR can be circumvented using quantum gravity \c{wj1963}. The very first regular BH (RBH) metric without a singularity was proposed by Bardeen in his seminal work \c{bj1968}. In \c{nedb}, authors have shown that a Bardeen BH can also be obtained from a specific form of NED. The formation of a de Sitter core prevents a singularity at the end of a massive object's gravitational collapse. Following the introduction of Bardeen BH, various RBH solutions were proposed \c{rbh1, rbh2, rbh3, rbh4}. RBHs are widely studied. Please see [\citenum{jos2011}-\citenum{rbh5}] for some of the recent studies in RBHs. \\

A central theme in recent studies is the investigation of plausible signatures of non-vacuum environments, such as DM halos, in astrophysical observations. To this end, there exist numerous phenomenological models [\citenum{kiselev}-\citenum{rayimbaev}] that are widely used to incorporate DM's effect. One such model is PFDM, in which the perfect-fluid nature of DM is assumed. This model is particularly helpful in explaining rotation curves in spiral galaxies \c{pf1}. Please see \c{pf2, pf3, pf4, pf5, pf6, pf7} for numerous studies in connection with PFDM. The minimal coupling between NED and DM provides a unique avenue to simultaneously probe them. Since the Bardeen BH may also result from a specific form of NED, DM's incorporation will provide a perfect opportunity to gauge the combined effect of NED and DM. With this in mind, authors in \c{pf6} obtained a static and spherically symmetric solution of a Bardeen BH surrounded by PFDM (BPFDM BH) and investigated its thermodynamic properties. Our aim is to expand this study to periodic orbits and gravitational waveforms emitted from extreme mass-ratio inspirals (EMRIs) where the central supermassive BH is modeled as a BPFDM BH.\\

In an EMRI, a stellar-mass object spirals inward towards an SMBH and, in the process, emits gravitational waves (GWs). Since the inspiral timescale spans several years, it provides a prolonged window of opportunity for precise measurements, as the signal-to-noise ratio is higher \c{emri1}. GWs emitted from EMRIs encode imprints of intrinsic features of the underlying spacetime and, as such, are potent tools to extract valuable information related to the SMBH and its surroundings. While spiraling down towards the SMBH, the stellar object traverses consecutive periodic orbits that serve as transition states \c{emri}. Periodic orbits are bound orbits where the particle returns to its initial position after some time, and as a result, the trajectory closes upon itself. For equatorial periodic orbits, a rational number $q$ is ascribed, and the $(z,w,v)$ taxonomy is widely used to organize them systematically \c{po1}. While the zoom number $z$ represents the number of highly elliptical trajectories the object travels before returning to its initial position, the whirl number $w$ is particularly important, representing the number of winds the object takes around the SMBH before zooming out. In zoom orbits, the stellar object travels in a weak gravitational field; hence, the motion is nearly Keplerian. In a whirling motion, the object moves near BH in a strong gravitational field and, as such, experiences strong acceleration. This variation in the gravitational field is clearly imprinted in GWs. Since the orbital dynamics and, consequently, the emitted GWs depend on the nature of the SMBH and are affected by the underlying environment, they provide an excellent avenue for probing our model and seeking signatures of NED and DM. Please see [\citenum{po2}-\citenum{po17}] for studies related to periodic orbits. \\

We organize the article as follows. Sec. I briefly introduces the BPFDM BH. Sec. II is where we study the timelike geodesics and Sec. III deals with periodic orbits. Sec. IV is where we study gravitational waveforms using the numerical "Kludge" method \c{kludge}. We conclude our article with Sec. V, where we provide a brief overview of results obtained. We use $\hbar=c=G=M=1$ throughout the article.
\section{bardeen bh surrounded by pfdm}
This manuscript entails probing a Bardeen BH surrounded by PFDM (BPFDM BH), which arises due to minimal coupling between NED and matter field, whose action is as follows:
\begin{widetext}
\begin{align}\label{action}
S=\int d^4x\sqrt{-g}\bigg[\frac{R}{2\kappa}-\f{2\mathcal{L}(F)}{\kappa}+\mathcal{L}_{dm}\bigg],
\end{align}
\end{widetext}
where $\kappa=8\pi G$, $R$ is the Ricci scalar, the NED Lagrangian for the Bardeen BH $\mathcal{L}(F)$ is \c{nedb}
\be
\mathcal{L}(F)=\frac{3M}{|g|^3}\lt(\frac{\sqrt{2g^2F}}{1+\sqrt{2g^2F}}\rt) \quad\text{with the field invariant}\quad F=\f{1}{4}F\mn F\MN,\label{lf}
\ee
$F\mn=\partial{\mu}A_{\nu}-\partial_{\nu}A_{\mu}$ being the NED field strength, and $\mathcal{L}_{dm}$ is the Lagrangian for the PFDM field. The NED field strength assumes the same form as in \c{nedb}. Parameters $g$ and $M$ in Eq. (\ref{lf}) are, respectively, associated with the magnetic charge and mass of the BH. Relevant field equations for the composite system are
\ba\nn
&&R_{\mu \nu }-\frac{1}{2}g_{\mu \nu }R=\kappa  T_{\mu\nu}^{\tx{DM}}+2\lt(\f{\pt \mc{L}}{\pt F}F_{\mu \lambda}F_{\nu}^{\lambda}-g_{\mu\nu}\mc{L}(F)\rt),\\\nn
&&\nabla_\mu\left(\frac{\partial \mathcal{L}(F)}{\partial F} F^{\nu \mu}\right)=0 \quad \text{and} \quad \nabla_\mu\left(* F^{\nu \mu}\right)=0,\label{fe}
\ea
where $*F^{\mu \nu}$ denotes the Hodge dual of $F^{\mu \nu}$ and the energy-momentum tensor of PFDM, $ T_{\mu\nu}^{\tx{DM}}$, is given by \c{kiselev}
\ba\nn
T_{\mu}^{\nu(\tx{DM})}&=&diag\lt[-\rho,p_r,p_\theta,p_\phi\rt],\\\nn
&=&diag\lt[\f{\b}{8\pi r^3}, \f{\b}{8\pi r^3}, -\f{\b}{16\pi r^3}, -\f{\b}{16\pi r^3} \rt].
\label{tdm}
\ea
We will assume $\b \leq 0$ throughout so that the weak energy condition $\rho \geq 0$ is satisfied. Solving field equations (\ref{fe}) results in the following static and spherically symmetric metric for BPFDM BH \c{pf6}:
\be
ds^2=-f(r)dt^2+f(r)^{-1}dr^2+r^2 d\theta^2+r^2 \sin^2\theta d\phi^2,\label{final}
\ee
with
\be
f(r)=1-\f{2Mr^2}{(r^2+g^2)^{\f{3}{2}}}+\f{\b}{r}ln\f{r}{|\b|}.
\ee
Limit $\b \rightarrow 0$ reverts the metric (\ref{final}) back to the Bardeen BH, whereas a further limit $g \rightarrow 0$ yields the metric for the \s BH. As shown in \c{pf6}, a BPFDM BH no longer remains a regular BH owing to the presence of PFDM background. Subsequent sections search for imprints of magnetic charge and DM in the orbital dynamics of particle motion around BPFDM BHs.
\section{effective potential governing particle motion in the background of bpfdm spacetime}
The effective potential governing the motion of test particles is central to the study of timelike geodesics and reveals significant features of the background spacetime. The Lagrangian for a free test particle, assuming its motion is confined to the equatorial plane, is
\ba\nn
\mathscr{L}&=&\f{1}{2}\lt(g_{tt} \dot{t}^2+g_{rr}\dot{r}^2+g_{\phi\phi} \dot{\phi}^2\rt),\\
&=&\f{1}{2}\lt(-f(r) \dot{t}^2+\f{\dot{r}^2}{f(r)}+r^2 \dot{\phi}^2\rt),\label{lagr} \ea
where $\dot{} \equiv \f{d}{d \tau}$, $\tau$ being the affine parameter. Since the Lagrangian bears no explicit dependence on time and azimuthal angle, test particles along timelike geodesics in the background of spacetime (\ref{final}) are endowed with two conserved quantities: namely, specific energy $E$ and specific angular momentum $L$ defined as:
\ba\nn
E&=&-p_{t}=-\f{\pt \mathscr{L}}{\pt \dot{t}}=f(r) \dot{t},\\\nn
L&=&p_{\phi}=\f{\pt \mathscr{L}}{\pt \dot{\phi}}=r^2 \dot{\phi}.\label{pt}
\ea
Above equations in tandem with the relation $u_{\mu}u^{\mu}=-1$ followed by the four-velocity $u^{\mu}=\f{d x^{\mu}}{d\tau}$ of test particles along timelike geodesics produce the following radial differential equation of motion:
\ba\nn
\dot{r}^2&=&E^2-f(r)\lt(1+\f{L^2}{r^2}\rt),\\
&=& E^2-V_{\text{eff}}(r),\label{radial}
\ea
where $V_{\text{eff}}(r)=f(r)\lt(1+\f{L^2}{r^2}\rt)$ is the effective potential that dictates particle dynamics. The magnetic charge and DM enter through $f(r)$. As evident from Eq. (\ref{radial}), particle motion is allowed for $E \geq V_{\text{eff}}(r)$, whereas extrema of $V_{\text{eff}}(r)$ locate bound orbits. The effective potential exhibits only one extremum when it satisfies $\f{\partial^2 \veff}{\partial r^2}=0$. This condition yields the location of the innermost stable circular orbit, and the corresponding angular momentum is $L_{ISCO}$. As the angular momentum increases, the effective potential displays two extrema: one for the unstable inner orbit and another for the stable outer orbit.
\begin{figure}[H]
\begin{center}
\begin{tabular}{cc}
\includegraphics[width=0.4\columnwidth]{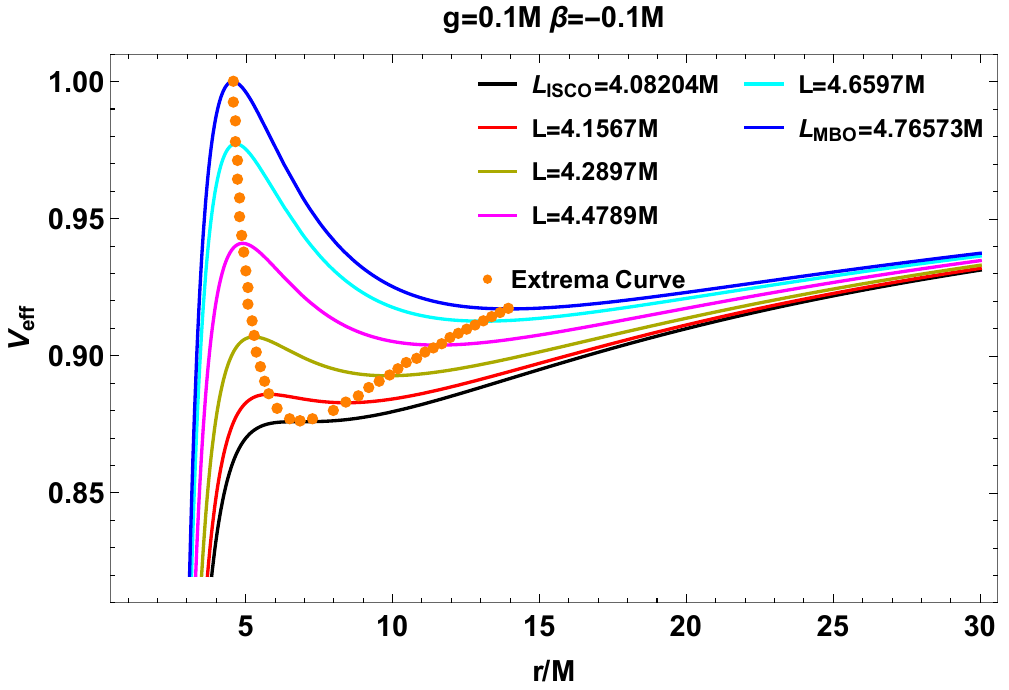}&
\includegraphics[width=0.4\columnwidth]{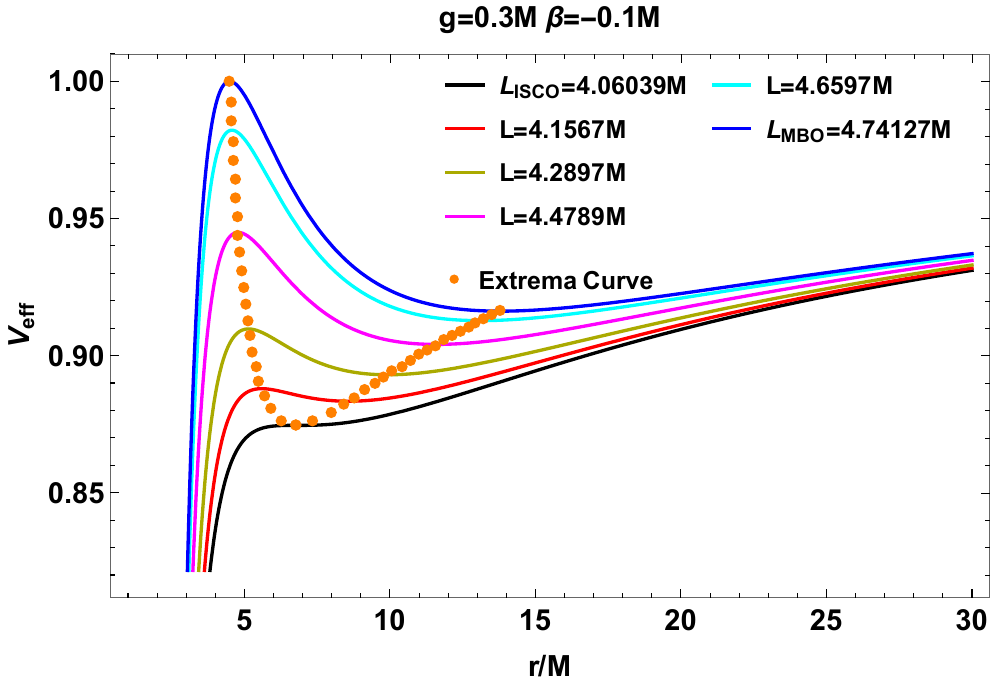}\\
\includegraphics[width=0.4\columnwidth]{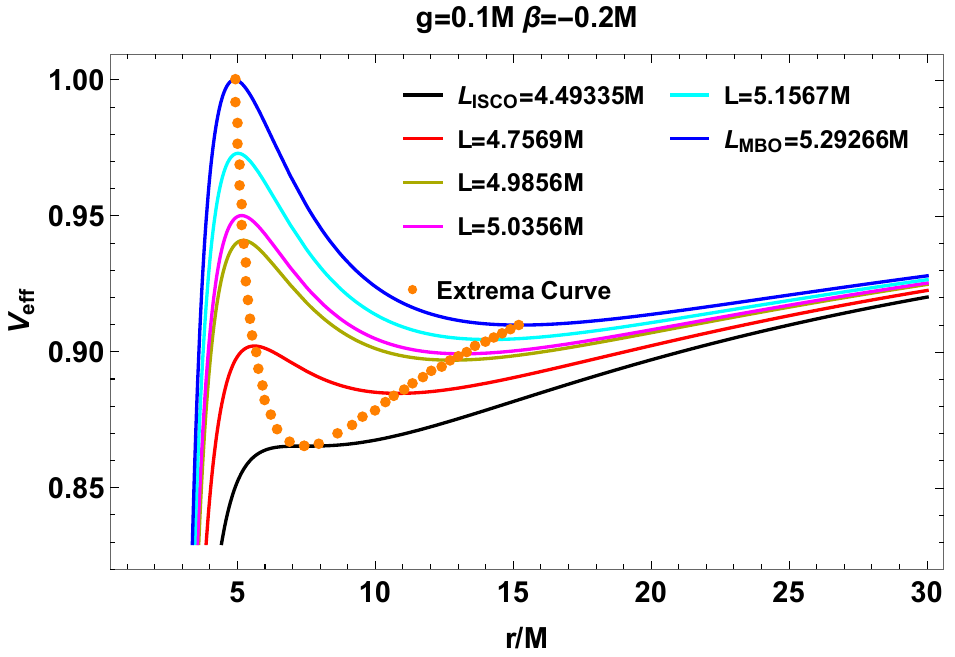}&
\includegraphics[width=0.4\columnwidth]{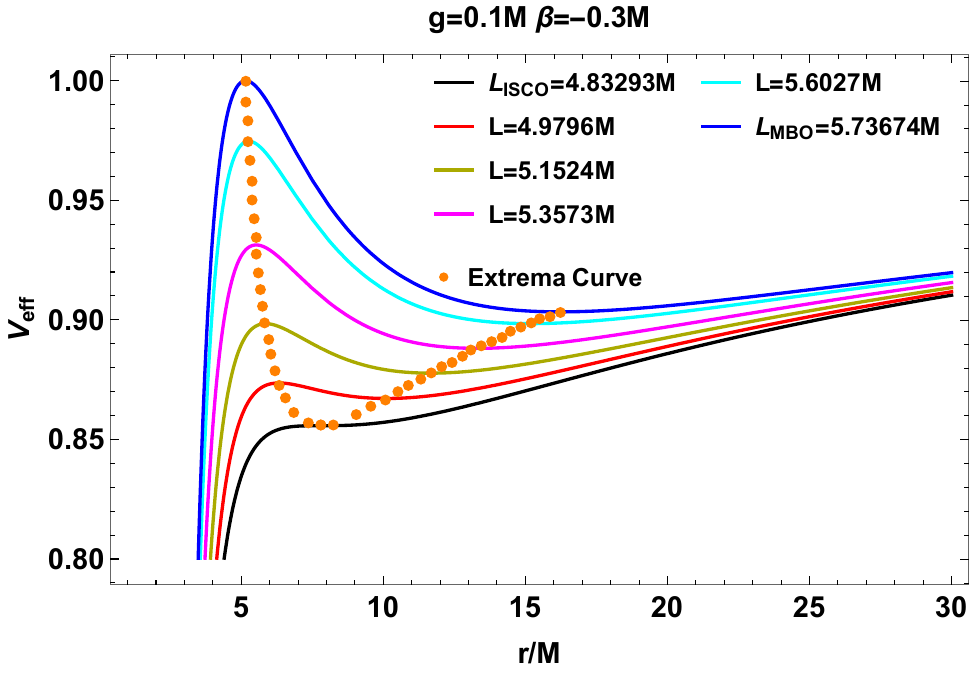}
\end{tabular}
\caption{The effective potential $\veff$ vs $r/M$ for different values of model parameters.}\label{veff}
\end{center}
\end{figure}
The extrema curve in Fig. (\ref{veff}) clearly demonstrates this fact. When the specific energy of test particles reaches $1$, we obtain a marginally bound orbit (MBO) whose corresponding specific angular momentum is $\lm$. Since $\veff \rightarrow 1$ for an asymptotic observer, test particles with $E>1$ exhibit unbounded orbits as $\dot{r}^2 >0$ in the limit $r \rightarrow \infty$ for them. As such, MBO marks the limiting case of the innermost bound orbit, and bound orbits are possible when $\ei\leq E \leq 1$. Particles with $E<\ei$ will eventually plunge into the BH, whereas particles with $E>1$ will be unbounded. By imposing suitable conditions on the effective potential and its derivatives, we obtain the locations and associated conserved quantities for MBO and ISCO.
\subsection{INNERMOST STABLE CIRCULAR ORBITS}
Following conditions, if imposed on $\veff$ and its first and second derivatives with respect to $r$, will yield position $\ri$, angular momentum $\li$, and energy $\ei$ coresponding to ISCO:
\begin{equation}\label{Isco}
V_{\mathrm{eff}}=E^2,\qquad \f{\partial \veff}{\partial r}=0, \qquad \text{and}\qquad \f{\partial^2 \veff}{\partial r^2}=0.
\end{equation}
ISCO's radius $\ri$ can be obtained by numerically solving the equation
\be
r_{\mathrm{ISCO}}=\frac{3 f\left(r_{\mathrm{ISCO}}\right) f^{\prime}\left(r_{\mathrm{ISCO}}\right)}{2 f^{\prime 2}\left(r_{\text {ISCO}}\right)-f\left(r_{\text {ISCO}}\right) f^{\prime \prime}\left(r_{\text {ISCO}}\right)},
\ee
whereas the corresponding conserved quantities are given by
\ba
L_{\mathrm{ISCO}}&=&r_{\mathrm{ISCO}}^{3 / 2} \sqrt{\frac{f^{\prime}\left(r_{\mathrm{ISCO}}\right)}{2 f\left(r_{\mathrm{ISCO}}\right)-r_{\mathrm{ISCO}} f^{\prime}\left(r_{\mathrm{ISCO}}\right)}},\\
E_{\mathrm{ISCO}}&=&\frac{f\left(r_{ISCO}\right)}{\sqrt{f\left(r_{\mathrm{ISCO}}\right)-r_{\mathrm{ISCO}} f^{\prime}\left(r_{\mathrm{ISCO}}\right)}}.
\ea
The prime above connotes differentiation with respect to the radial co-ordinate $r$.
\begin{figure}[H]
\begin{center}
\begin{tabular}{cc}
\includegraphics[width=0.4\columnwidth]{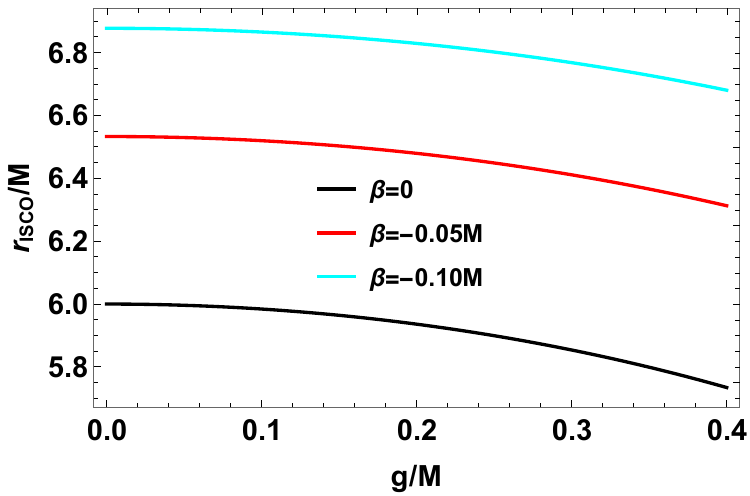}&
\includegraphics[width=0.4\columnwidth]{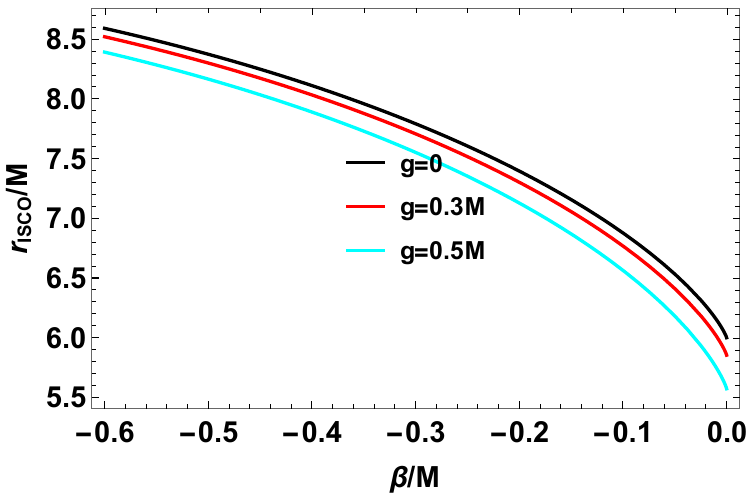}
\end{tabular}
\caption{Variation of ISCO radius $\ri$ with magetic charge $g$ (left panel) and DM parameter $\b$ (right panel).}\label{risco}
\end{center}
\end{figure}
\begin{figure}[H]
\begin{center}
\begin{tabular}{cc}
\includegraphics[width=0.4\columnwidth]{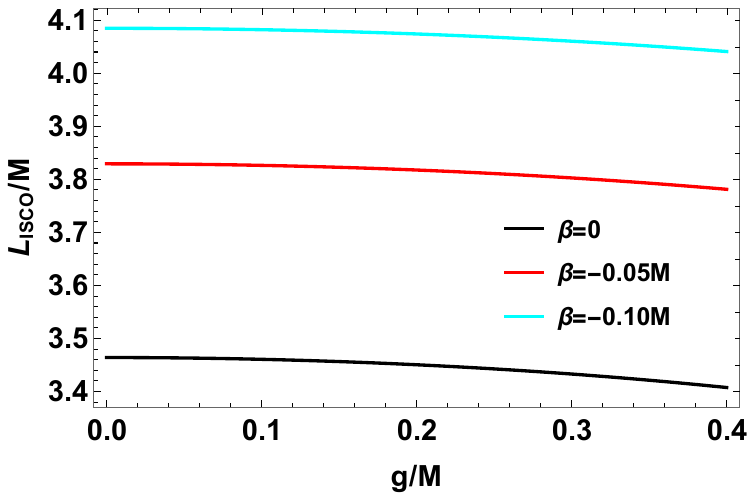}&
\includegraphics[width=0.4\columnwidth]{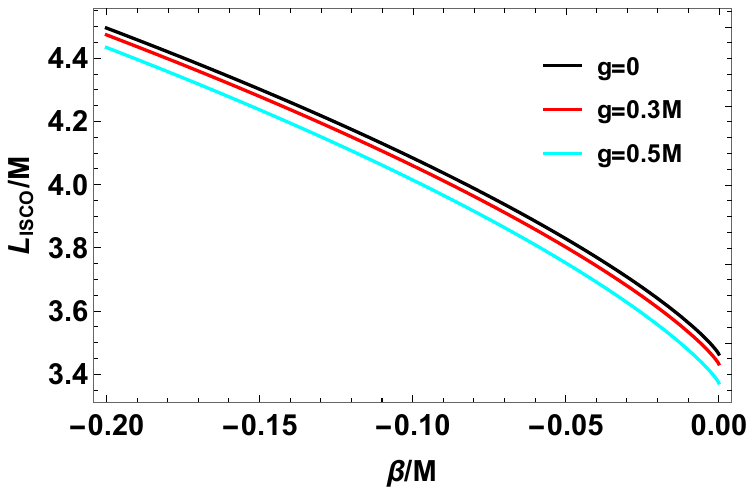}
\end{tabular}
\caption{Variation of angular momentum $\li$ with magetic charge $g$ (left panel) and DM parameter $\b$ (right panel).}\label{lisco}
\end{center}
\end{figure}
\begin{figure}[H]
\begin{center}
\begin{tabular}{cc}
\includegraphics[width=0.4\columnwidth]{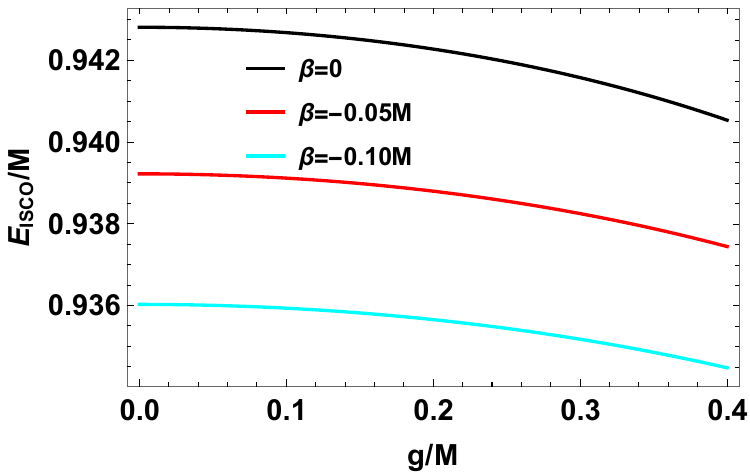}&
\includegraphics[width=0.4\columnwidth]{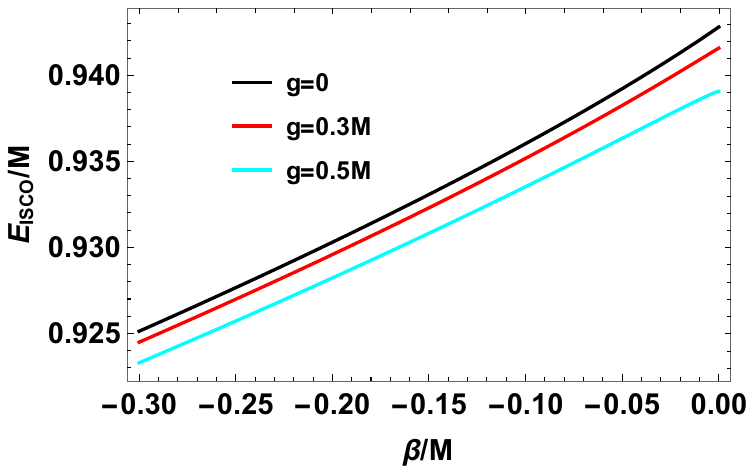}
\end{tabular}
\caption{Variation of energy $\ei$ with magetic charge $g$ (left panel) and DM parameter $\b$ (right panel).}\label{eisco}
\end{center}
\end{figure}
Fig. (\ref{risco}), (\ref{lisco}), and (\ref{eisco}) graphically elucidate how parameters $g$ and $\b$ influence $\ri$, $\li$, and $\ei$. As these figures show, increasing magnetic charge adversely affects these quantities. While increasing $\b$ adversely impacts ISCO radius and angular momentum, its effect on $\ei$ is favorable. The change in ISCO-related quantities is found to be steeper for a change in the DM parameter than for a similar change in magnetic charge. These findings suggest that for a BPFDM BH with non-zero values of magnetic charge and DM parameter, an ISCO exists at a smaller radius with smaller angular momentum and energy than for a \s BH, implying significant impact of magnetic charge and DM on particle dynamics.
\subsection{MARGINALLY BOUND ORBITS}
For marginally bound orbits, the energy reaches its maximum allowable value for a bound orbit, i.e., 1. The conditions to be imposed on the effective potential are:
\begin{equation}
\veff=1\qquad\text{and}\qquad\f{\partial \veff}{\partial r}=0.
\end{equation}
The above conditions lead to the following equation for the MBO radius $\rm$:
\begin{equation}
r_{\mathrm{MBO}} =\frac{2 f\left(r_{\mathrm{MBO}}\right)\left(1-f\left(r_{\mathrm{MBO}}\right)\right)}{f^{\prime}\left(r_{\mathrm{MBO}}\right)}.
\end{equation}
It is numerically solved to locate MBO. The corresponding angular momentum is
\be
L_{\mathrm{MBO}} =\sqrt{\frac{1-f\left(r_{\mathrm{MBO}}\right)}{f\left(r_{\mathrm{MBO}}\right)}} r_{\mathrm{MBO}}.
\ee
\begin{figure}[H]
\begin{center}
\begin{tabular}{cc}
\includegraphics[width=0.4\columnwidth]{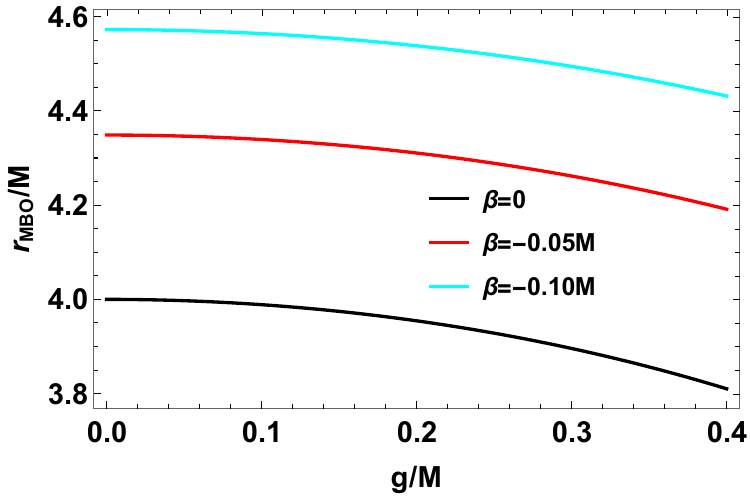}&
\includegraphics[width=0.4\columnwidth]{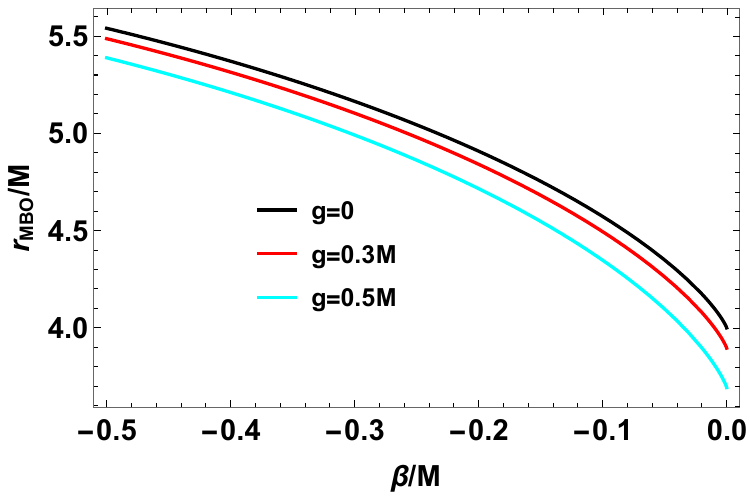}
\end{tabular}
\caption{Variation of MBO radius $\rm$ with magetic charge $g$ (left panel) and DM parameter $\b$ (right panel).}\label{rmbo}
\end{center}
\end{figure}
\begin{figure}[H]
\begin{center}
\begin{tabular}{cc}
\includegraphics[width=0.4\columnwidth]{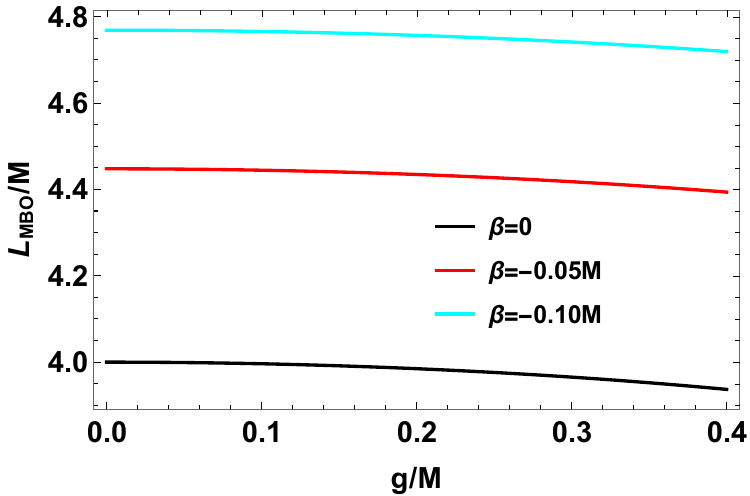}&
\includegraphics[width=0.4\columnwidth]{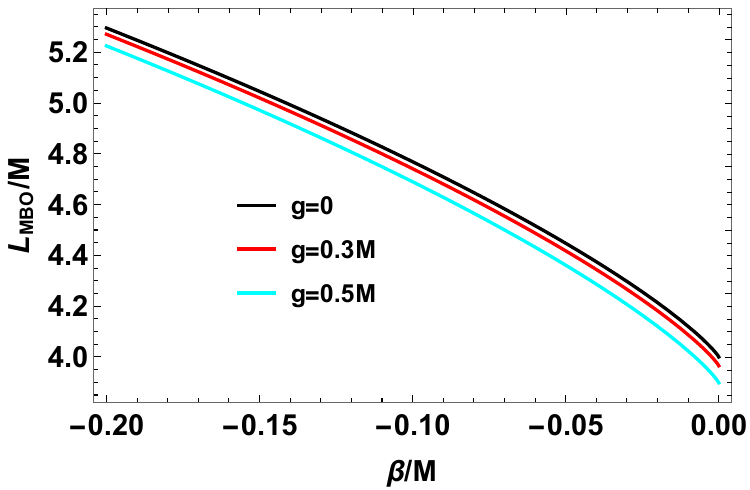}
\end{tabular}
\caption{Variation of angular momentum $\lm$ with magetic charge $g$ (left panel) and DM parameter $\b$ (right panel).}\label{lmbo}
\end{center}
\end{figure}
Similar to ISCO, Figs. (\ref{rmbo}) and (\ref{lmbo}) exhibit the adverse impact of magnetic charge and DM on MBO radius and angular momentum, thereby implying the existence of MBOs at lower radii and with smaller angular momenta in the case of BPFDM BHs than \s BHs. Our findings here and in the previous subsection clearly demonstrate the significant impact of model parameters on related quantities of bound orbits.\\
\begin{figure}[H]
\begin{center}
\begin{tabular}{cc}
\includegraphics[width=0.4\columnwidth]{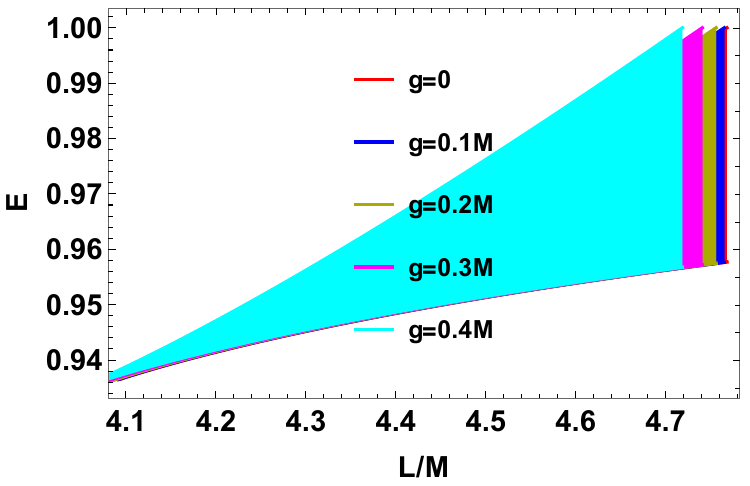}&
\includegraphics[width=0.4\columnwidth]{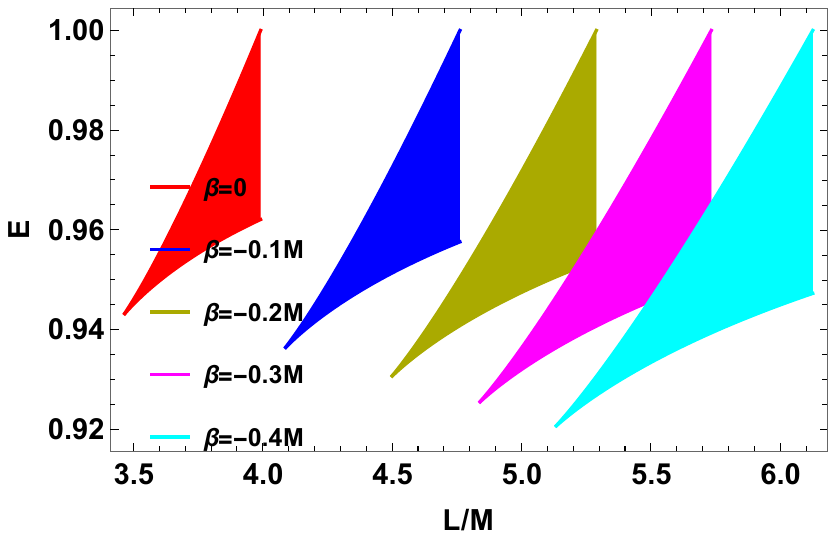}
\end{tabular}
\caption{Parameter space for timelike bound orbits in $(E,\,L/M)$ plane. The left one displays the allowed region's variation with $g$ keeping $\b$ fixed at $-0.1M$, whereas the right one demonstrates its variation with $\b$ keeping $g$ fixed at $0.15M$. }\label{para}
\end{center}
\end{figure}
For different values of $g$ and $\b$, Fig. (\ref{para}) illustrates parameter space for timelike bound orbits in $(E,\,L/M)$ plane. It is bounded below by ISCO and above by MBO. The presence of DM significantly alters the permissible region, allowing wider ranges of energy and angular momentum for the formation of bound orbits, with allowed angular momenta shifting towards higher values. The imprints of magnetic charge on the allowed parameter space are less pronounced. In this case, the permissible region shrinks as $g$ increases, with a smaller impact on energy than on angular momentum.
\section{Periodic Orbits}
Periodic orbits are a special type of bound orbits where the test particle returns to its initial position after a finite time, and the resultant trajectory, as a result, closes upon itself. For such orbits, frequencies in $\phi$- and $r$-directions is a rational number $q$ defined as \c{po1}:
\begin{equation}\label{q}
q=\frac{\omega_{\phi}}{\omega_{r}}-1=\frac{\Delta\phi}{2\pi}-1,
\end{equation}
where the azimuthal angle accumulated over one radial oscillation, $\Delta\phi$, is defined as
\begin{equation}\label{dp}
{{\Delta\phi=\oint\mathrm{d}\phi}}\\ {{{}=2\int_{r_{p}}^{r_{a}}\frac{\dot{\phi}}{\dot{r}}\mathrm{d}r}}\\ {{{}=2\int_{r_{p}}^{r_{a}}{\frac{L}{r^2\sqrt{E^2-f(r)(1+\frac{L^2}{r^2})}}}}}.
\end{equation}
The factor $2$ arises because, in one radial period, a test particle travels from apoapsis $r_a$ to periapsis $r_p$ and back to apoapsis. Eq. (\ref{q}), together with Eq. (\ref{dp}), shows that the rational number $q$ depends on the energy $E$, angular momentum $L$, and the function $f(r)$. Since the angular momentum for bound orbits varies from $\li$ to $\lm$, we may parametrize $L$ as:
\begin{equation}\label{ang}
L=L_{\mathrm{ISCO}}+\epsilon(L_{\mathrm{MBO}}-L_{\mathrm{ISCO}}),
\end{equation}
where $\epsilon \in [0,\,1]$, with $0$ and $1$ giving the angular momenta for ISCO and MBO, respectively. For $\epsilon > 1$, particle trajectories become unbounded. \\
Following \c{po1}, periodic orbits are characterized by three integers-$(z,w,v)$. $z$ is the zoom number that a periodic orbit traces before returning to its initial position. $w$ is the whirl number representing the number of whirls a test particle takes around the central object before going back to apastron. The final integer, $v$, is the vertex number signifying the vertex hit by the particle after the initial apastron. These integers together provide a taxonomy for periodic orbits. The rational number, $q$, is associated with these integers through the relation \c{po1}:
\begin{equation}
q=w+\frac{v}{z}.
\end{equation}
\begin{figure}[H]
\begin{center}
\begin{tabular}{cc}
\includegraphics[width=0.4\columnwidth]{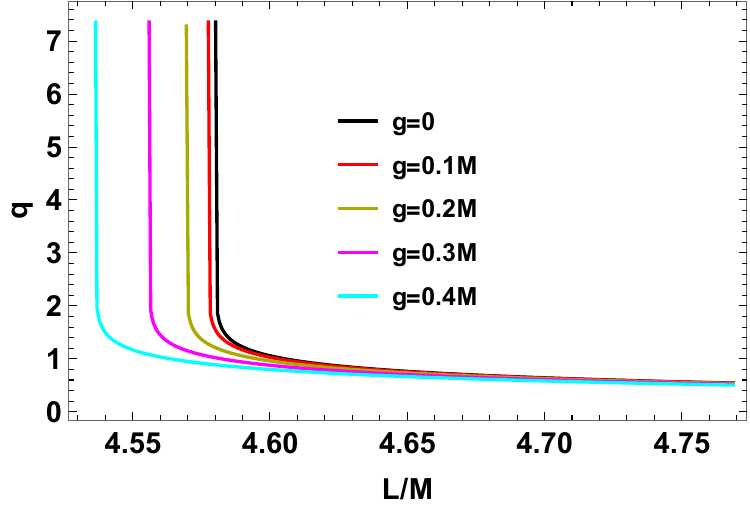}&
\includegraphics[width=0.4\columnwidth]{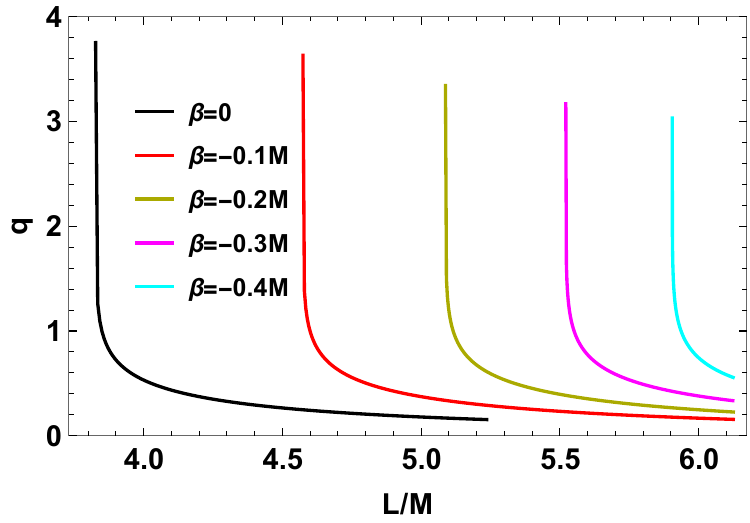}
\end{tabular}
\caption{Rational number $q$ as a function of $L$. The left panel is for different values of $g$ keeping $\b=-0.1M$, and the right one is for different values of $\b$ keeping $g=0.15M$. For both these cases, energy is kept at $0.98$. }\label{ql}
\end{center}
\end{figure}
Fig. (\ref{ql}) illustrates the behavior of the rational number $q$ as a function of angular momentum for different values of $g$ and $\b$ at a fixed energy. The rational number exhibits divergence as the angular momentum approaches its lower bound, signifying the onset of extreme whirling motion of particles around the BH before returning to apastron. As the angular momentum decreases, the whirling motion gets suppressed, and for large $L$, trajectories become nearly Keplerian. As evident from Fig. (\ref{ql}), the motion becomes whirl-dominated at lower angular momentum for higher values of $g$ or $\b$.
\begin{figure}[H]
\begin{center}
\begin{tabular}{cc}
\includegraphics[width=0.4\columnwidth]{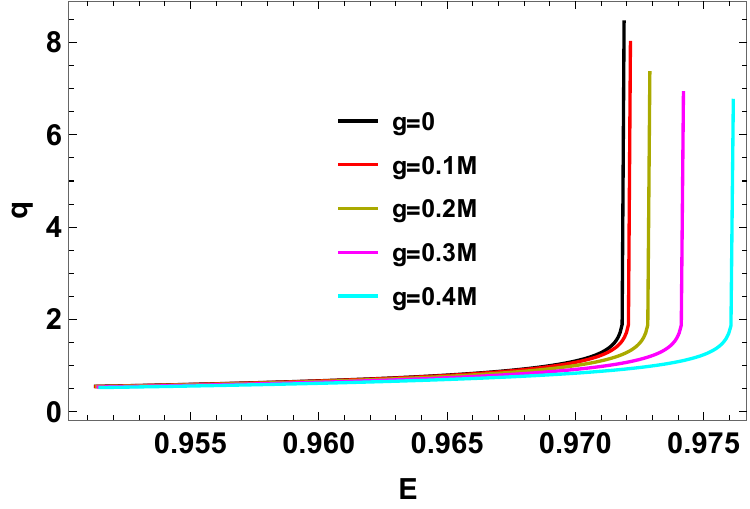}&
\includegraphics[width=0.4\columnwidth]{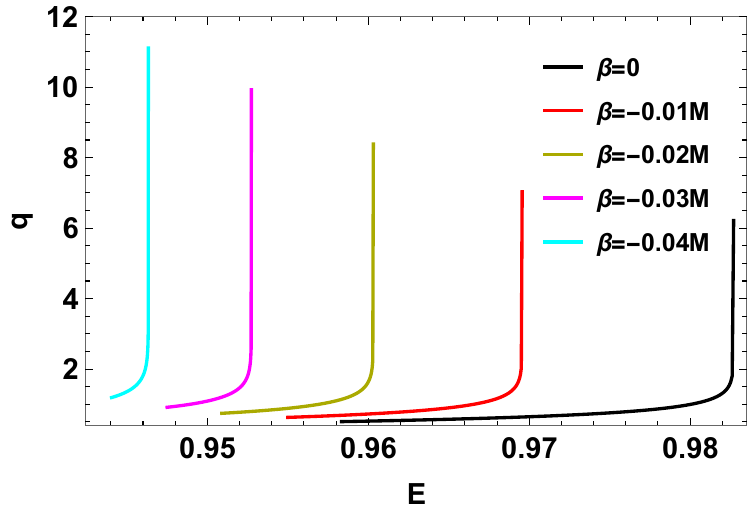}
\end{tabular}
\caption{Rational number $q$ as a function of $E$. The left panel is for different values of $g$ keeping $\b=-0.1M$, $L=4.5$, and the right one is for different values of $\b$ keeping $g=0.15M$, $L=3.85$. }\label{qe}
\end{center}
\end{figure}
Fig. (\ref{ql}) demonstrates how $q$ varies with energy $E$ for different values of magnetic charge and DM parameter at a fixed value of angular momentum. As the energy increases, the whirling motion dominates, leading to higher values of $q$. As $E$ approaches its upper bound, $q$ diverges owing to extreme whirling motion. The extreme zoom-whirl motion sets in at higher energies when we increase either the magnetic charge or the DM parameter. \\
\begin{table}[H]
\begin{center}
\vspace{-0.2cm}
\begin{tabular}{cccccccccc}
\hline
$g/M$ & $L/M$ & $E_{(1,1,0)}$ & $E_{(1,2,0)}$ & $E_{(2,1,1)}$ & $E_{(2,2,1)}$ & $E_{(3,1,2)}$ & $E_{(3,2,2)}$ & $E_{(4,1,3)}$ & $E_{(4,2,3)}$ \\
\hline
$0.$ & $4.42672$ & $0.961645$ & $0.964804$ & $0.964317$ & $0.964736$ & $0.964521$ & $0.96474$ & $0.964582$ & $0.964741$ \\
$0.1$ & $4.42389$ & $0.961577$ & $0.964634$ & $0.964265$ & $0.964687$ & $0.96447$ & $0.964691$ & $0.964532$ & $0.964693$ \\
$0.2$ & $4.41532$ & $0.96137$ & $0.964484$ & $0.964106$ & $0.964511$ & $0.964316$ & $0.964544$ & $0.96438$ & $0.964545$ \\
$0.3$ & $4.40083$ & $0.961013$ & $0.964229$ & $0.963835$ & $0.964287$ & $0.964053$ & $0.964292$ & $0.96412$ & $0.964293$ \\
$0.4$ & $4.38004$ & $0.960487$ & $0.963857$ & $0.963437$ & $0.963919$ & $0.963669$ & $0.963925$ & $0.96374$ & $0.963926$ \\
\hline
\end{tabular}
\end{center}
\caption{Energy $E$ associated with different q-periodic orbits. Here, the magnetic charge is varied keeping DM parameter fixed at $\b=-0.1M$. \label{eg}}
\end{table}
Table (\ref{eg}) and (\ref{eb}) list energies for different periodic orbits with varied $(z,w,v)$ configurations keeping $\epsilon=0.5$.
\begin{table}[H]
\begin{center}
\vspace{-0.2cm}
\begin{tabular}{cccccccccc}
\hline
$\b/M$ & $L/M$ & $E_{(1,1,0)}$ & $E_{(1,2,0)}$ & $E_{(2,1,1)}$ & $E_{(2,2,1)}$ & $E_{(3,1,2)}$ & $E_{(3,2,2)}$ & $E_{(4,1,3)}$ & $E_{(4,2,3)}$ \\
\hline
$-0.4$ & $5.62523$ & $0.952633$ & $0.956087$ & $0.955662$ & $0.95615$ & $0.955898$ & $0.956155$ & $0.955969$ & $0.956157$ \\
$-0.3$ & $5.28187$ & $0.955375$ & $0.958684$ & $0.95828$ & $0.958743$ & $0.958504$ & $0.958748$ & $0.958572$ & $0.958749$ \\
$-0.2$ & $4.88982$ & $0.958294$ & $0.961476$ & $0.96109$ & $0.961532$ & $0.961305$ & $0.961537$ & $0.96137$ & $0.961538$ \\
$-0.1$ & $4.42033$ & $0.961491$ & $0.964572$ & $0.964199$ & $0.964626$ & $0.964406$ & $0.96463$ & $0.964469$ & $0.964632$ \\
$0$ & $3.72399$ & $0.965205$ & $0.968224$ & $0.967858$ & $0.968277$ & $0.968061$ & $0.968281$ & $0.968123$ & $0.968283$ \\
\hline
\end{tabular}
\end{center}
\caption{Energy $E$ associated with different q-periodic orbits. Here, the DM parameter is varied, keeping magnetic charge fixed at $g=0.15M$. \label{eb}}
\end{table}
In Table (\ref{eg}), the magnetic charge varies from $0$ to $0.4M$, keeping the DM parameter fixed at $-0.1M$. In Table (\ref{eb}), the DM parameter varies from $-0.4M$ to $0$, keeping magnetic charge fixed at $0.15M$. The angular momentum and the energy decrease with increasing $g$ or $\b$.\\
\begin{figure*}[htbp]
\center{
\subfigure[E=0.961577]{\label{Periodicorbits6a}
\includegraphics[width=5.4cm]{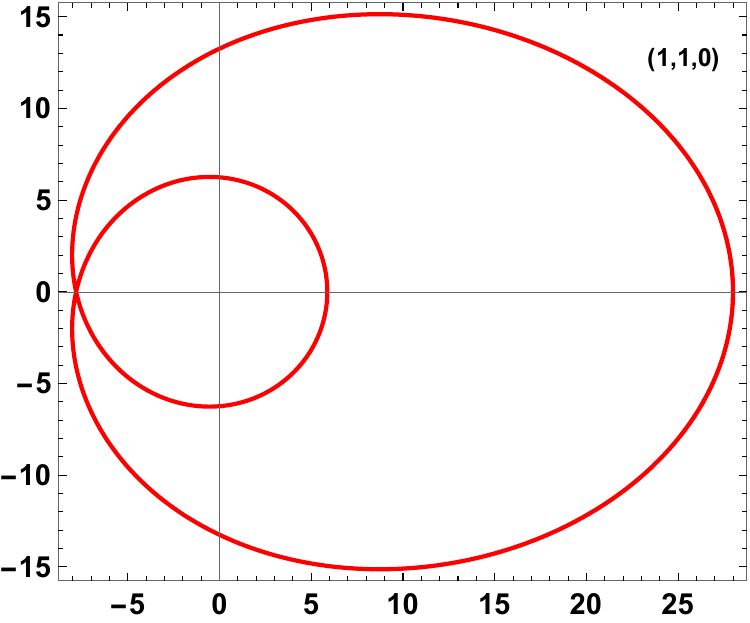}}
\subfigure[E=0.964634]{\label{Periodicorbits6b}
\includegraphics[width=5.1cm]{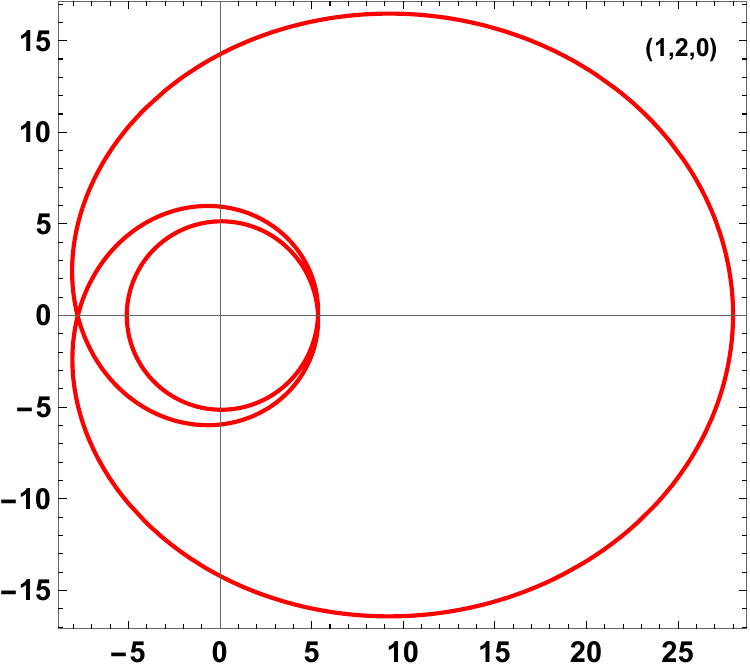}}
\subfigure[E=0.964695]{\label{Periodicorbits6c}
\includegraphics[width=5.1cm]{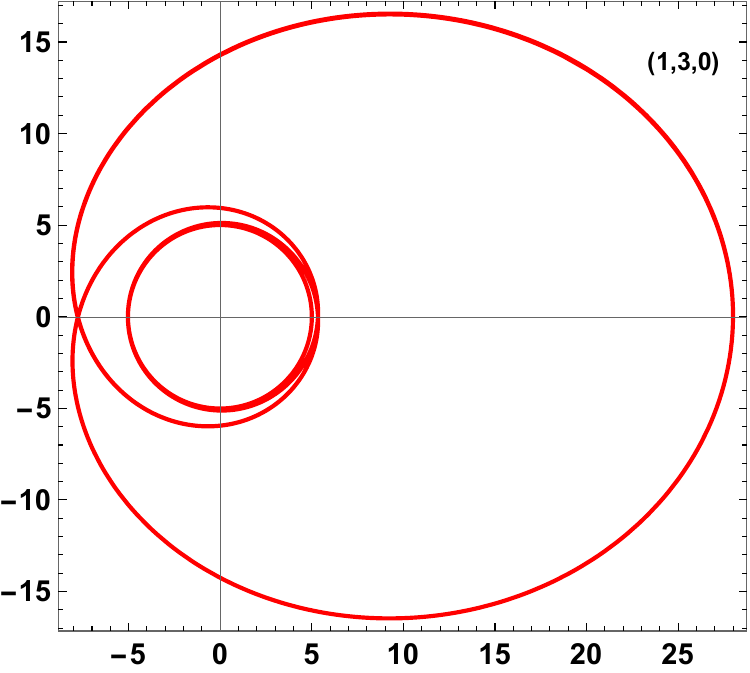}}\\
\subfigure[E=0.964265]{\label{tPeriodicorbits6d}
\includegraphics[width=5.2cm, height=3.5cm]{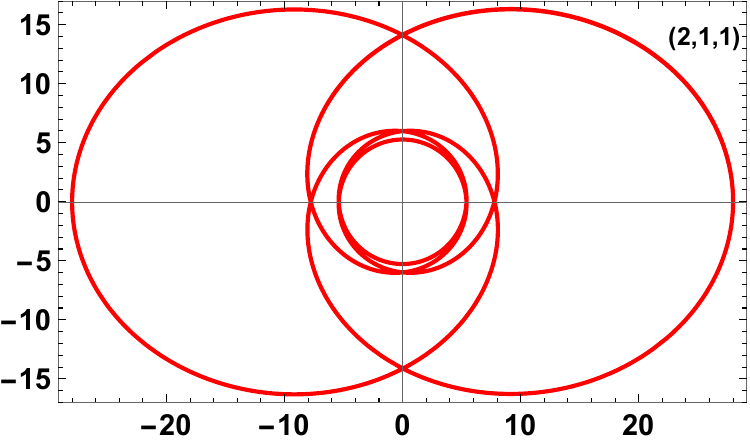}}
\subfigure[E=0.964687]{\label{Periodicorbits6e}
\includegraphics[width=5.1cm, height=3.5cm]{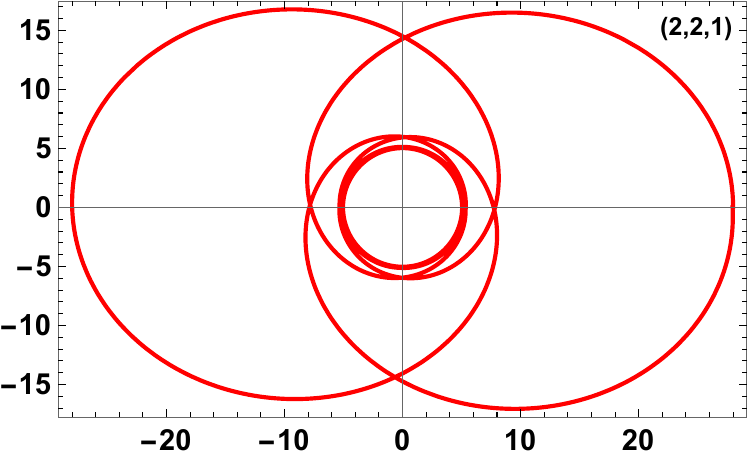}}
\subfigure[E=0.9646961]{\label{Periodicorbits6f}
\includegraphics[width=5.1cm, height=3.5cm]{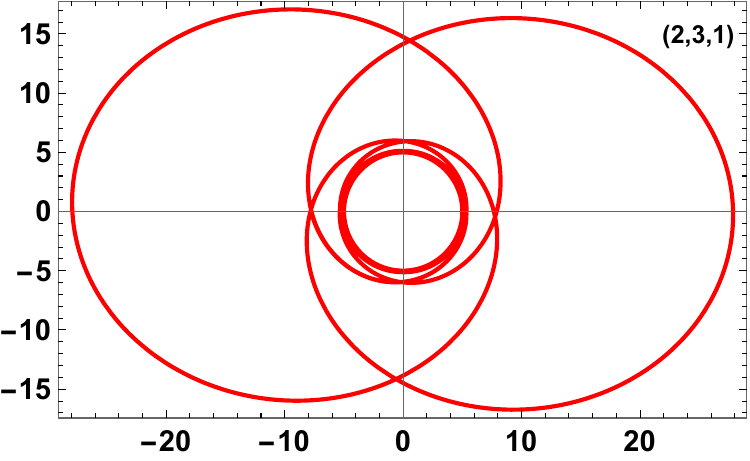}}\\
\subfigure[E=0.96447]{\label{Periodicorbits6g}
\includegraphics[width=5.1cm]{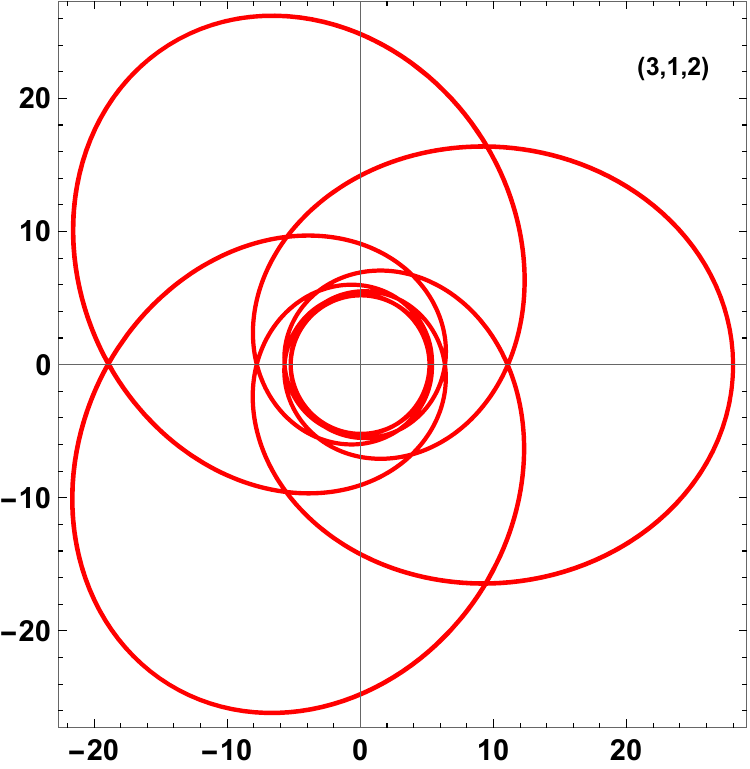}}
\subfigure[E=0.964691]{\label{Periodicorbits6h}
\includegraphics[width=5.1cm]{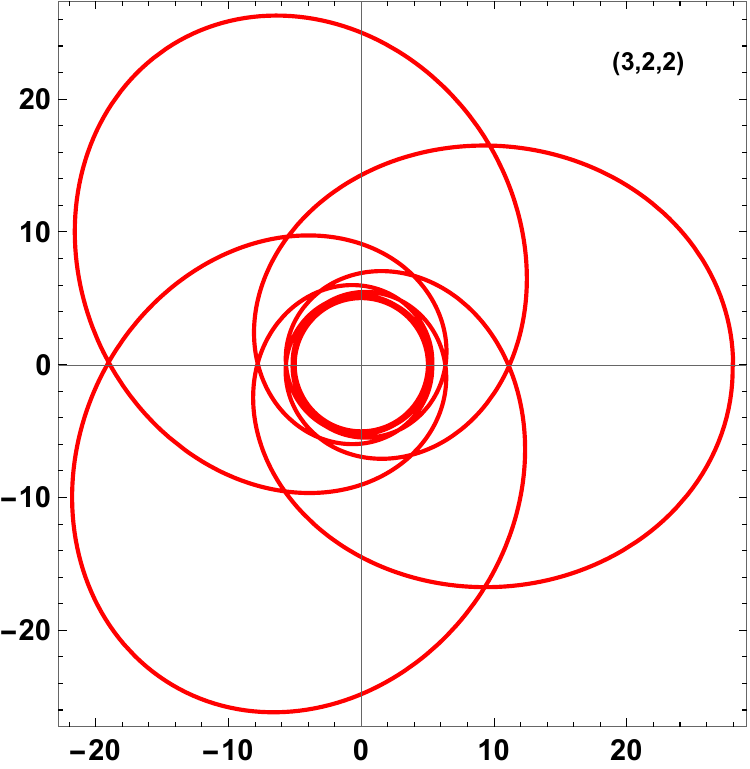}}
\subfigure[E=.964696]{\label{Periodicorbits6i}
\includegraphics[width=5.1cm]{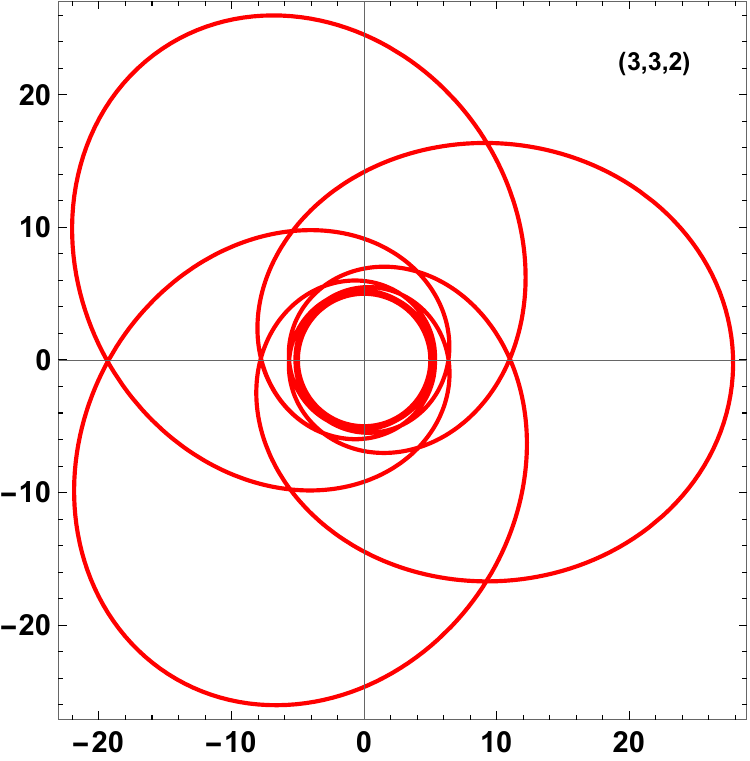}}\\
\subfigure[E=0.955848]{\label{tPeriodicorbits6j}
\includegraphics[width=5.1cm]{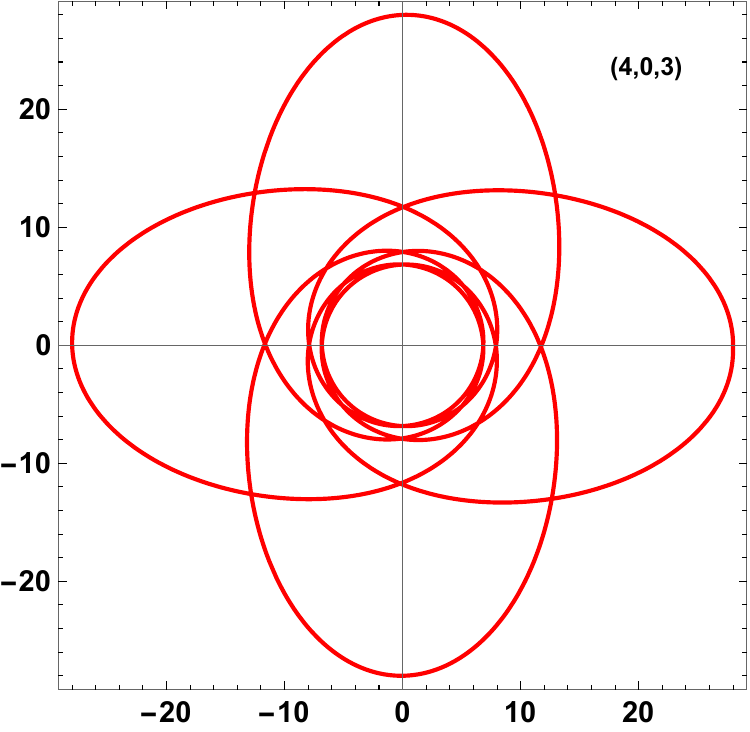}}
\subfigure[E=0.964532]{\label{Periodicorbits6k}
\includegraphics[width=5.1cm]{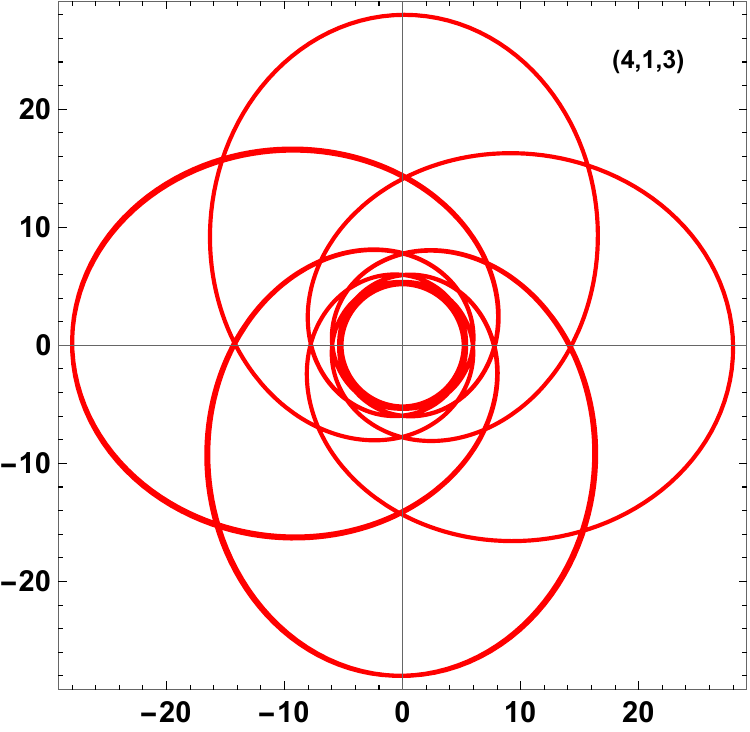}}
\subfigure[E=0.964693]{\label{Periodicorbits6l}
\includegraphics[width=5.1cm]{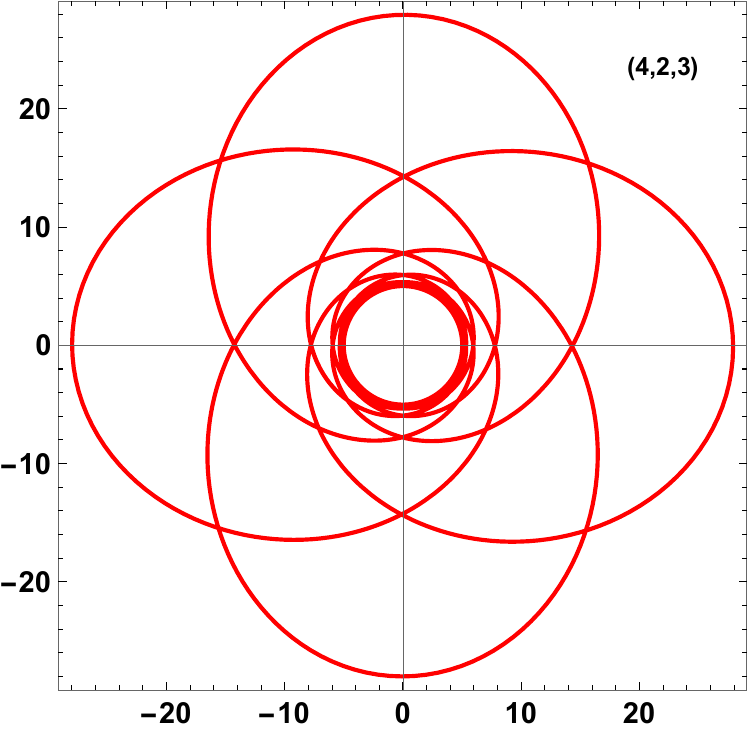}}
}
\captionsetup{justification=raggedright}
\caption{Periodic orbits for different $(z, w, v)$ configurations around the BPFDM BH with $g=0.1M$, $\b=-0.1M$, and $\epsilon=0.5$.}\label{tra}
\end{figure*}
\begin{figure*}[htbp]
\center{
\subfigure[L=4.40798]{\label{Periodicorbits6a}
\includegraphics[width=5.1cm]{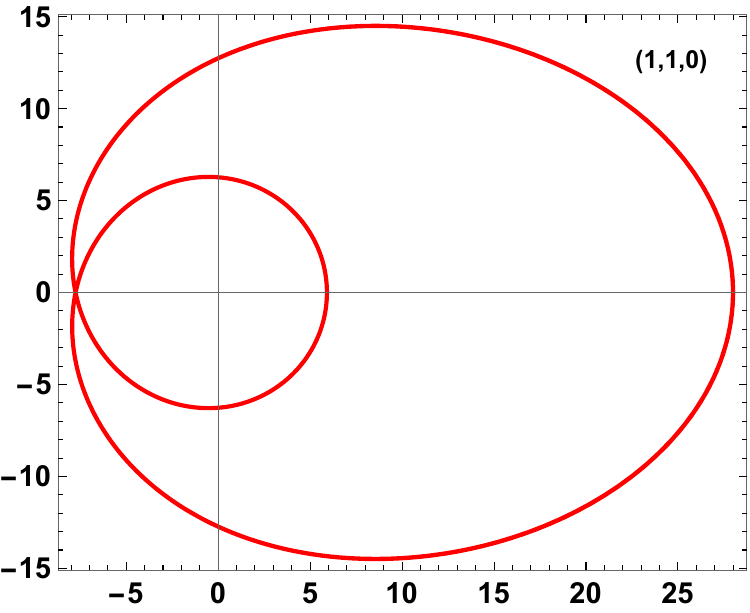}}
\subfigure[L=4.37512]{\label{Periodicorbits6b}
\includegraphics[width=5.1cm]{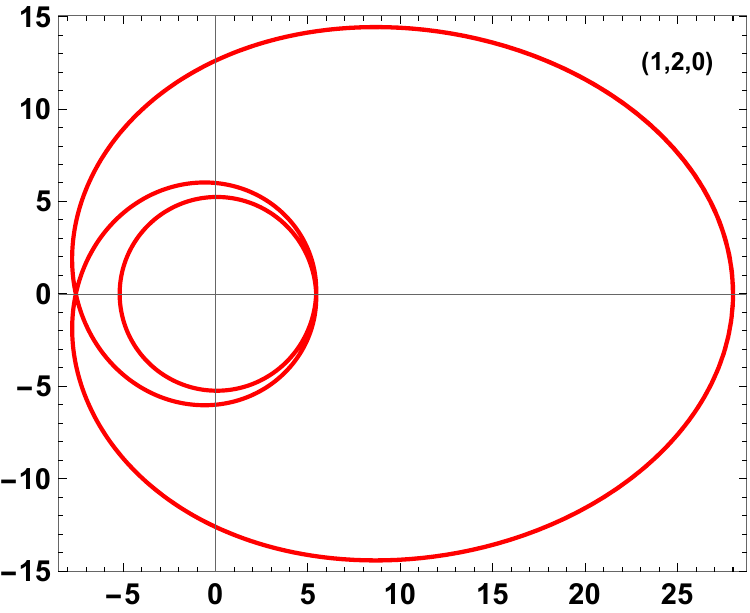}}
\subfigure[L=4.37435]{\label{Periodicorbits6c}
\includegraphics[width=5.1cm]{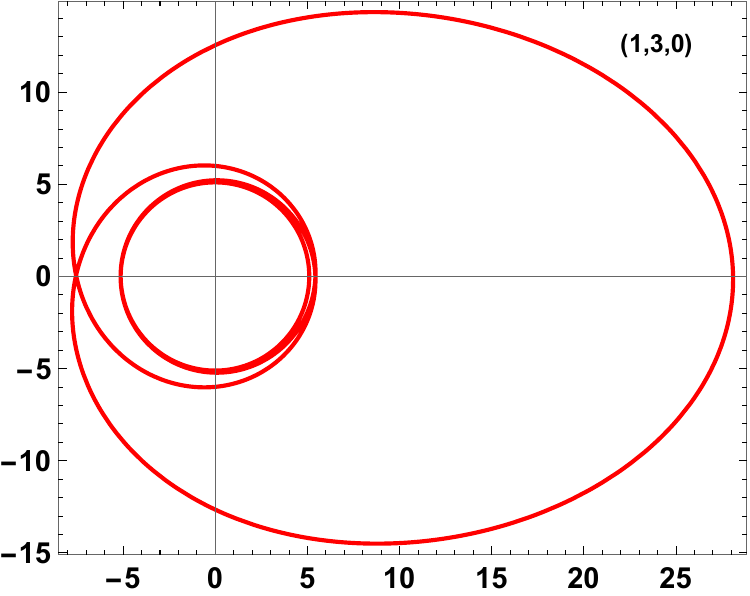}}\\
\subfigure[L=4.37949]{\label{tPeriodicorbits6d}
\includegraphics[width=5.2cm, height=3.5cm]{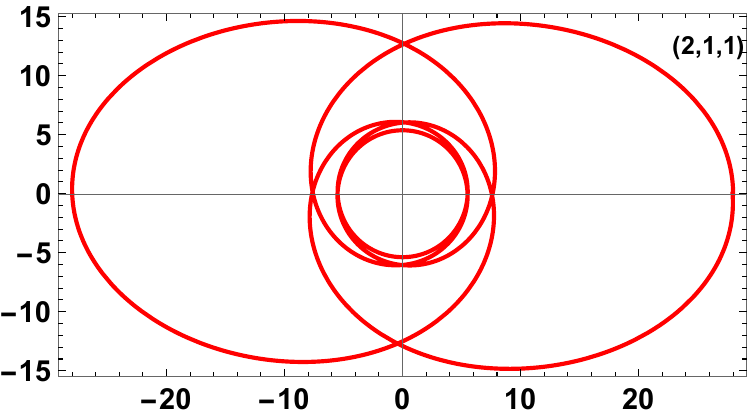}}
\subfigure[L=4.37445]{\label{Periodicorbits6e}
\includegraphics[width=5.2cm, height=3.5cm]{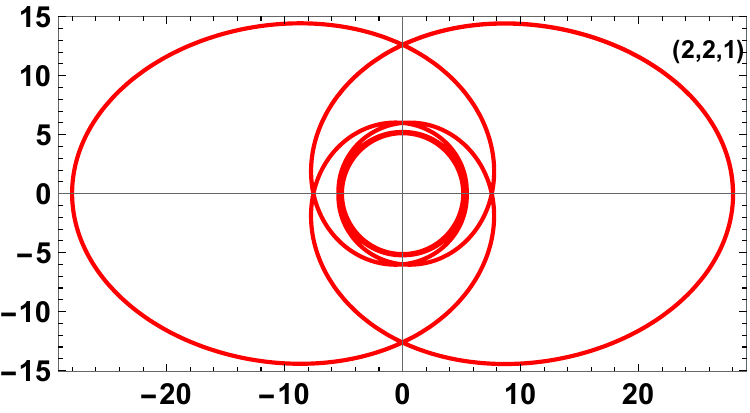}}
\subfigure[L=4.37433]{\label{Periodicorbits6f}
\includegraphics[width=5.25cm, height=3.5cm]{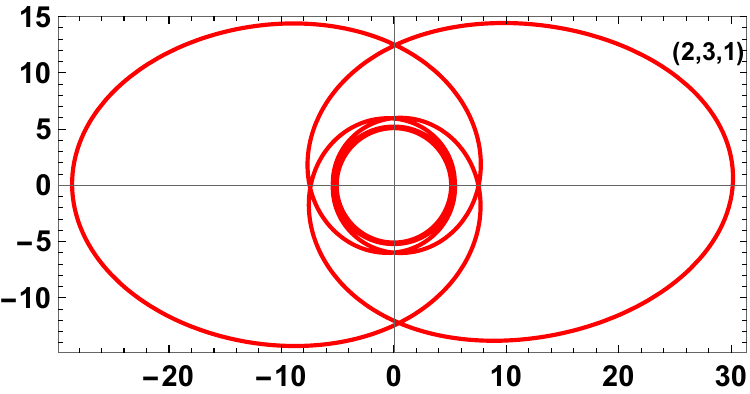}}\\
\subfigure[L=4.37707]{\label{Periodicorbits6g}
\includegraphics[width=5.1cm]{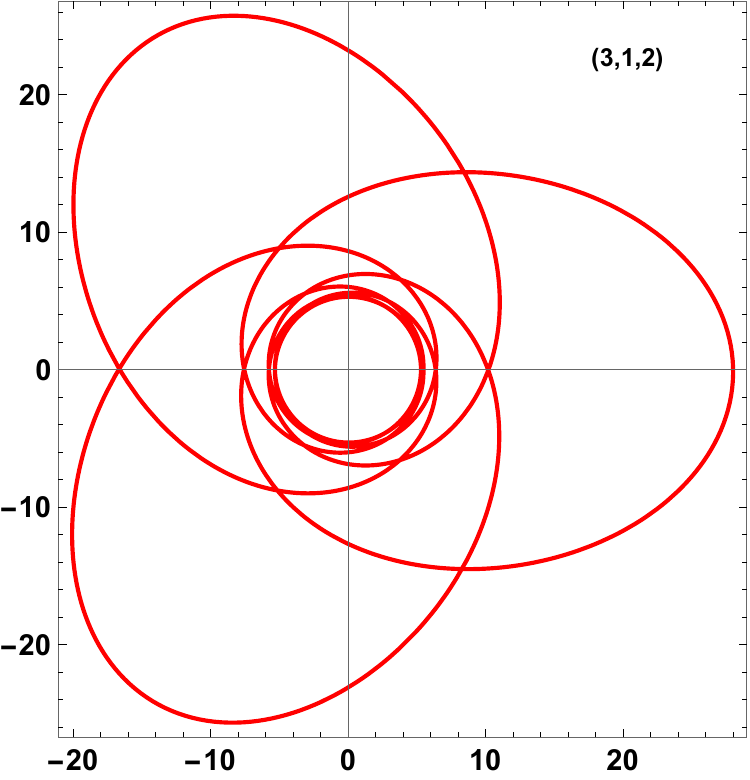}}
\subfigure[L=4.3744]{\label{Periodicorbits6h}
\includegraphics[width=5.1cm]{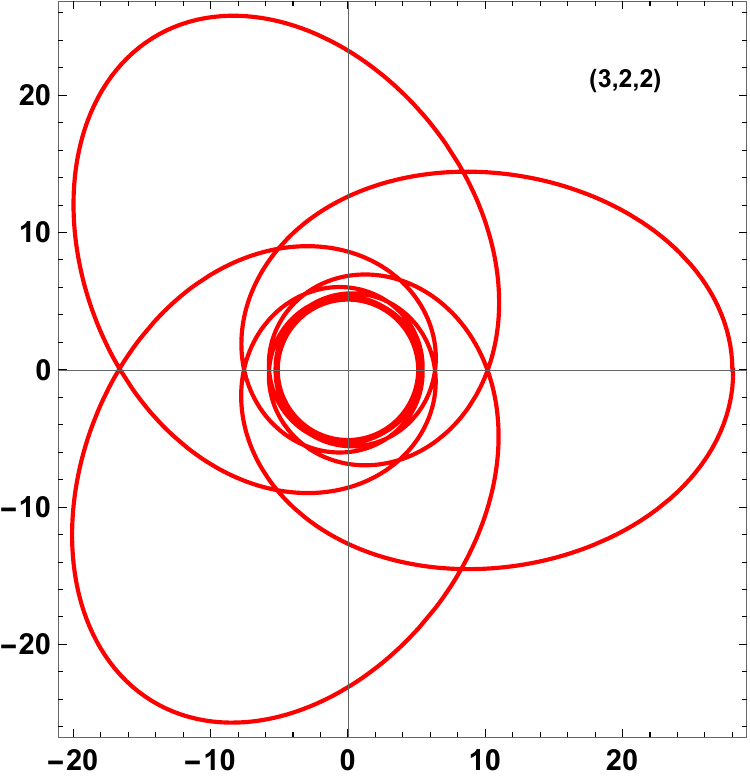}}
\subfigure[L=4.37433]{\label{Periodicorbits6i}
\includegraphics[width=5.1cm]{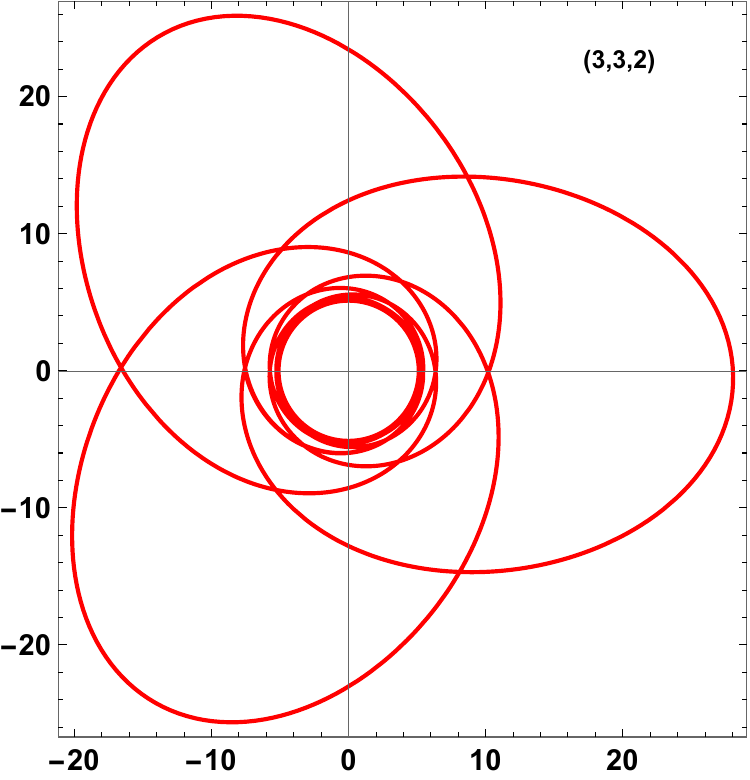}}\\
\subfigure[L=4.46401]{\label{tPeriodicorbits6j}
\includegraphics[width=5.1cm]{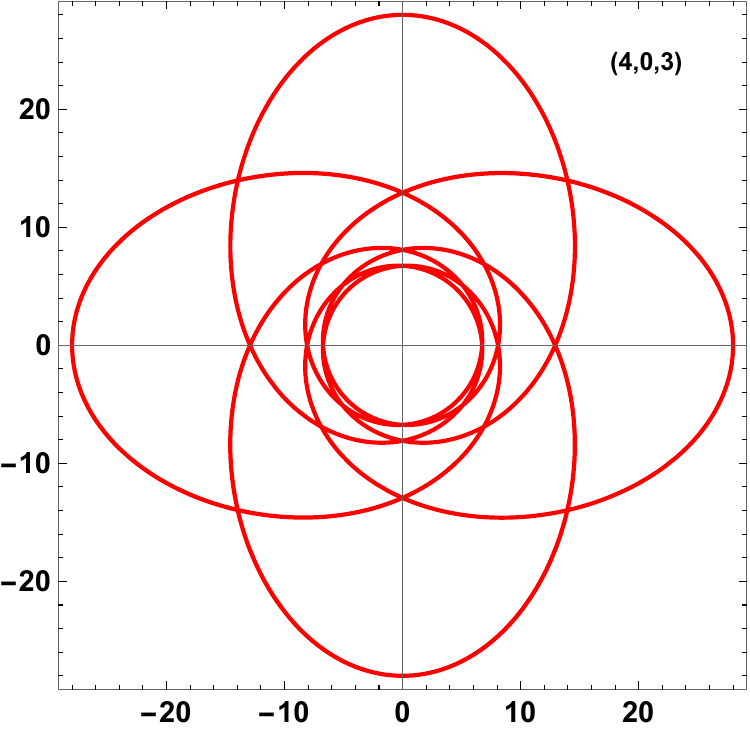}}
\subfigure[L=4.37634]{\label{Periodicorbits6k}
\includegraphics[width=5.1cm]{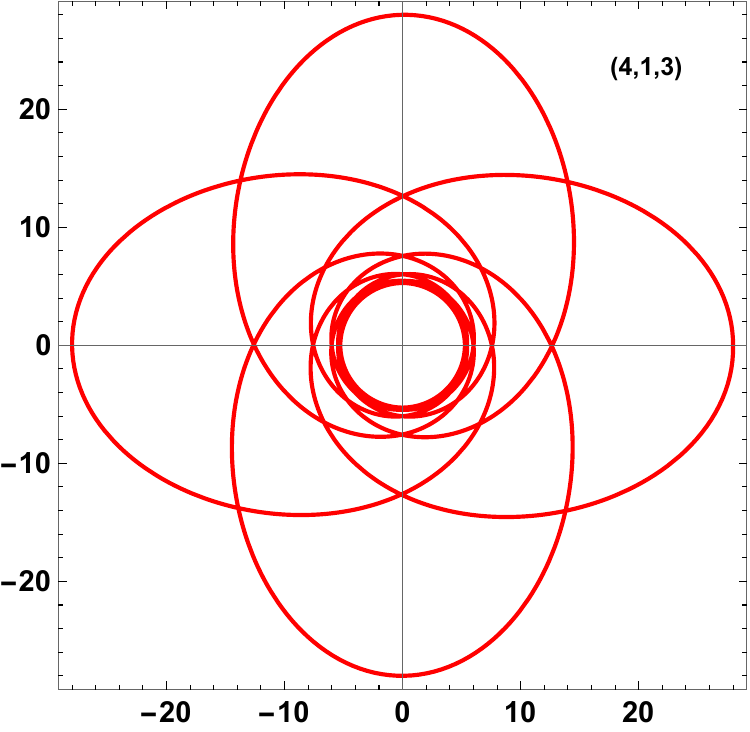}}
\subfigure[L=4.37438]{\label{Periodicorbits6l}
\includegraphics[width=5.1cm]{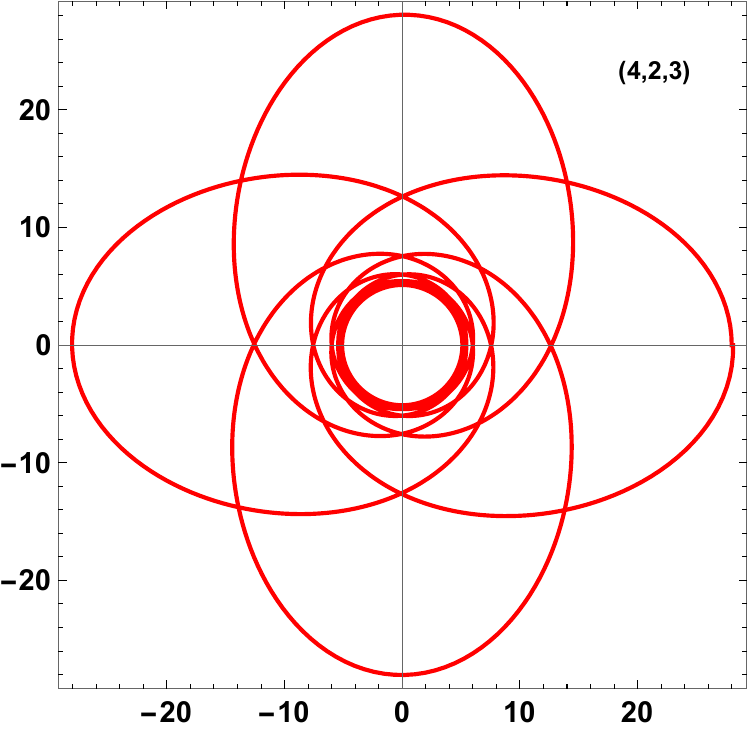}}
}
\captionsetup{justification=raggedright}
\caption{Periodic orbits for different $(z, w, v)$ configurations around the BPFDM BH with $g=0.1M$, $\b=-0.1M$, and $E=0.96$.}\label{tra1}
\end{figure*}
Fig. (\ref{tra}) and (\ref{tra1}) display periodic orbits for varied $(z, w, v)$ configurations. As the zoom number $z$ increases, the periodic orbits exhibit richer structural complexity, and particles zoom out a larger number of times before returning to their initial position. Increasing whirl number $w$, on the other hand, leads particles to wind a higher number of times around the central BH before closure, hence staying for a prolonged time in the strong-field region near the BH. It requires larger energy for the onset of whirl-dominated motion, as evident from Fig. (\ref{tra}), whereas Fig. (\ref{tra1}) showcases the fact that the whirling motion dominates when the angular momentum is smaller.
\section{GRAVITATIONAL WAVE FROM PERIODIC ORBITS}
\begin{figure*}[htbp]
\centering
\begin{minipage}{0.3\textwidth}
\centering
\includegraphics[width=\linewidth]{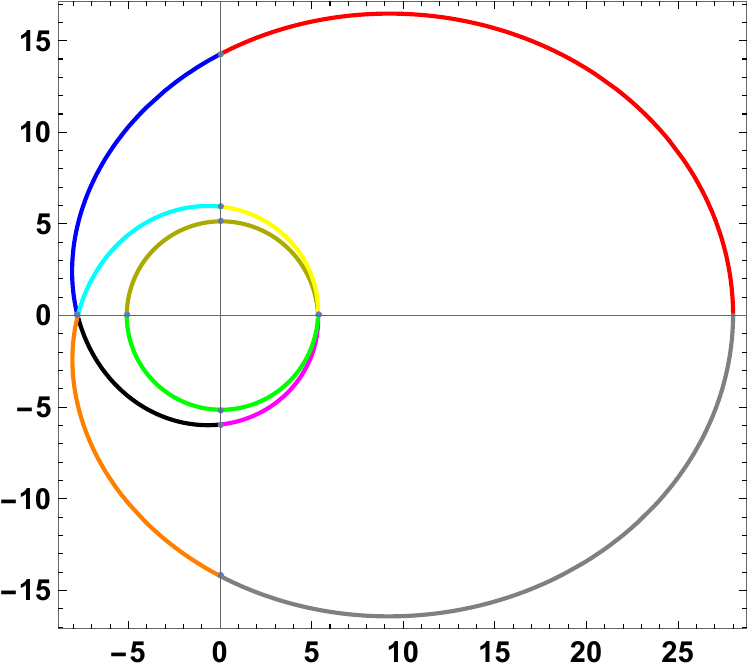}
(a) Periodic orbit.
\end{minipage}
\begin{minipage}{0.4\textwidth}
\centering
\includegraphics[width=\linewidth]{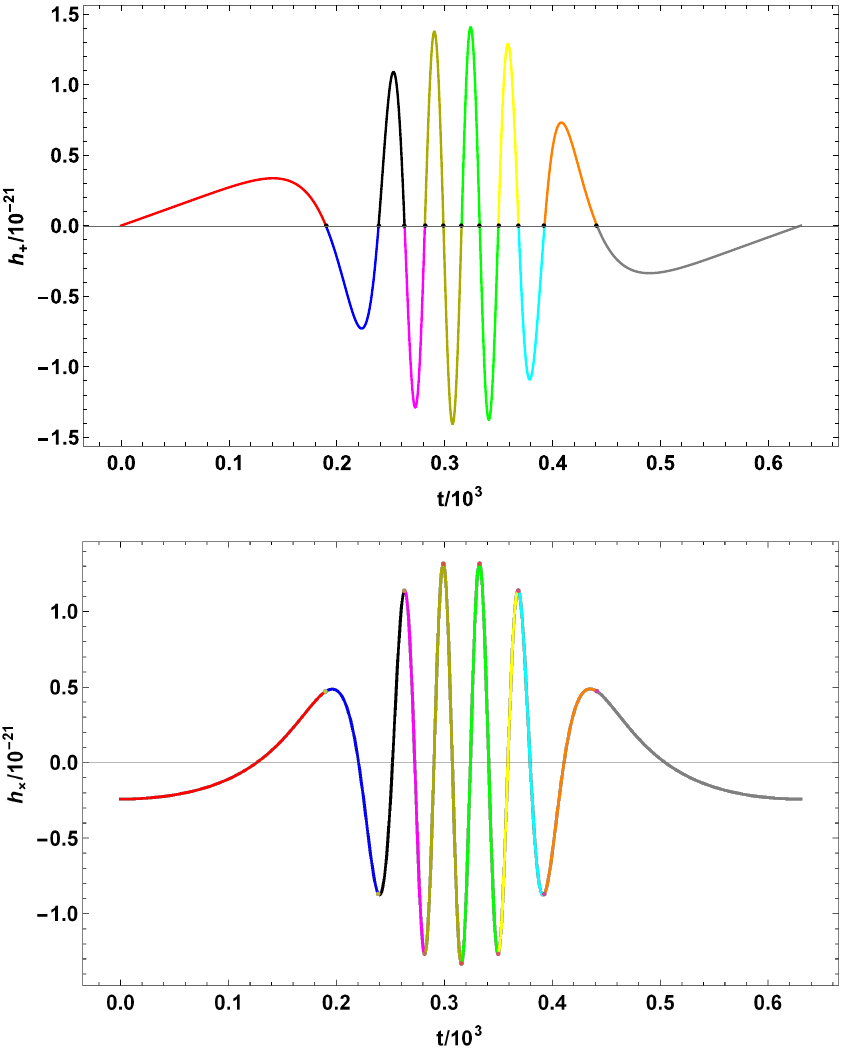}
(b) Gravitational waveforms.
\end{minipage}
\captionsetup{justification=raggedright}
\caption{Gravitational waveform for the periodic orbit with $(z,\,w,\,v)=(1,2,0)$. The left panel shows the periodic orbit, and the right panels show gravitational waveforms for polarizations $h_+$ and $h_\times$. Different colors are for different portions of the periodic orbit. We have taken $g=0.1M$ and $\b=-0.1M$.}
\label{120}
\end{figure*}

\begin{figure*}[htbp]
\centering
\begin{minipage}{0.3\textwidth}
\centering
\includegraphics[width=\linewidth]{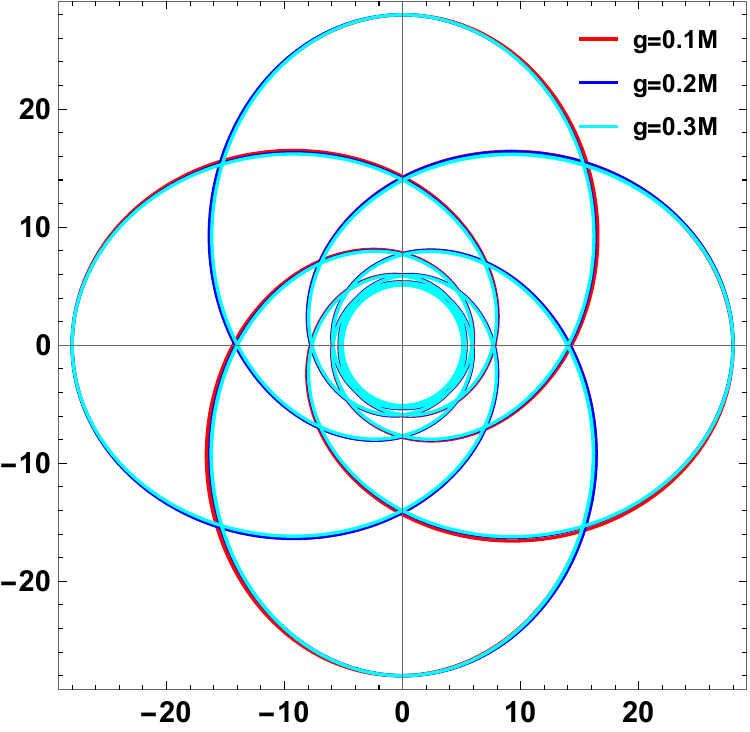}
(a) Periodic orbit.
\end{minipage}
\begin{minipage}{0.4\textwidth}
\centering
\includegraphics[width=\linewidth]{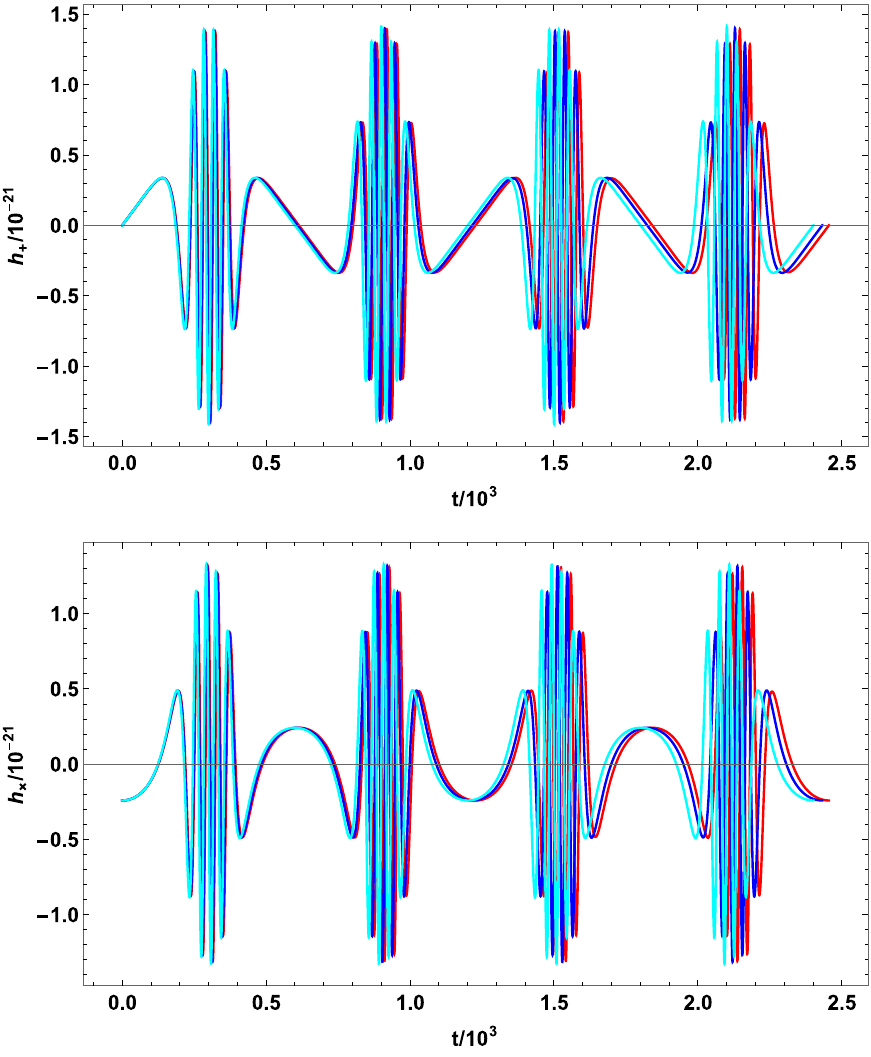}
(b) Gravitational waveforms.
\end{minipage}
\captionsetup{justification=raggedright}
\caption{Periodic orbits and gravitational waveforms for different values of magnetic charge with $(z,\,w,\,v)=(4,1,3)$. The left panel shows the periodic orbit, and the right panels show gravitational waveforms for polarizations $h_+$ and $h_\times$. We have taken $\b=-0.1M$.}
\label{413a}
\end{figure*}

\begin{figure*}[htbp]
\centering
\begin{minipage}{0.3\textwidth}
\centering
\includegraphics[width=\linewidth]{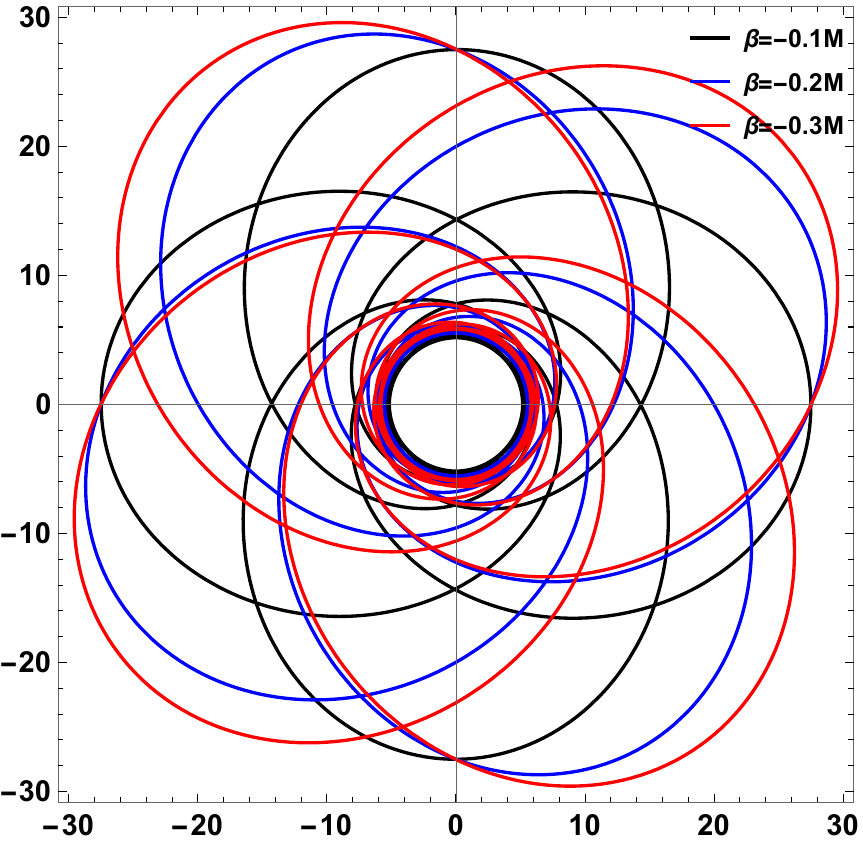}
(a) Periodic orbit.
\end{minipage}
\begin{minipage}{0.4\textwidth}
\centering
\includegraphics[width=\linewidth]{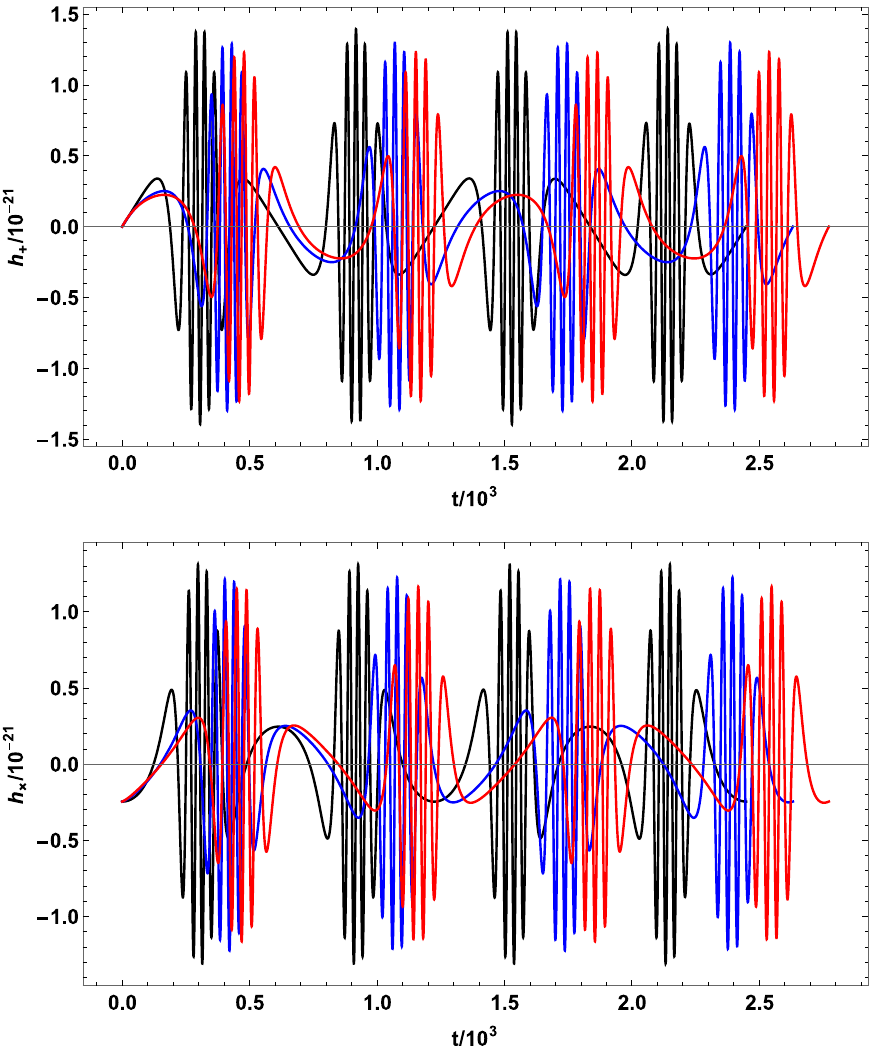}
(b) Gravitational waveforms.
\end{minipage}
\captionsetup{justification=raggedright}
\caption{Periodic orbits and gravitational waveforms for different values of DM parameter with $(z,\,w,\,v)=(4,1,3)$. The left panel shows the periodic orbit, and the right panels show gravitational waveforms for polarizations $h_+$ and $h_\times$. We have taken $\b=-0.1M$.}
\label{413b}
\end{figure*}

This section deals with the emission of GWs from periodic orbits around BPFDM BHs. In an EMRI, a smaller compact object follows a periodic orbit around a BPFDM BH. Since the mass of the orbiting object is much smaller than that of the central BH, we may neglect the backreaction of the smaller object on the geometry. The smaller object oscillates between the apoapsis and periapsis while in a periodic orbit and, in the process, winds several times in the strong field region near the event horizon of the BH. This results in bursts of GW and energy loss for the orbiting object. As a result, GWs emitted from EMRIs provide an excellent avenue to probe near-horizon geometry and glean invaluable information in connection with the intrinsic nature of the underlying spacetime.\\
In EMRIs, the compact object loses energy via GW emission and eventually inspirals toward the supermassive BH. Since the loss of energy and angular momentum due to GW emission is negligible over one orbital period, the adiabatic approximation [\citenum{adia1} - \citenum{adia7}] is apt for such a situation. Under this approximation, the energy and angular momentum of the smaller object are assumed to be fixed, allowing us to employ the geodesic equations to describe its motion. We utilize the numerical "Kludge" method \c{kludge} to generate the GW form. The "Kludge" method entails obtaining a particle's trajectory in a pseudo-flat spacetime, which is then used in the quadrupole formula to derive the gravitational waveform. \\
The trajectory of the smaller object is first obtained in the Boyer-Lindquist coordinates $(r,\,\theta,\,\phi)$, treated as hypothetical spherical coordinates, and is then projected onto a Cartesian frame $(x,\,y,\,z)$ where \c{kludge}
\begin{equation}\label{cart}
x=r \sin \theta \cos \phi, \quad y=r \sin \theta \sin \phi, \quad z=r \cos \theta .
\end{equation}
The linearized approximation in the weak-field limit, conjoined with the gauge condition, yields the following field equation.
\begin{equation}\label{linear}
\square\bar{h}^{\mu\nu}=-16\pi\mathcal{T}^{\mu\nu},
\end{equation}
where $\square$ is the wave operator in flat space and $\mathcal{T}^{\mu\nu}_{,\nu}=0$. The solution of the Eq. (\ref{linear}), in the slow-motion limit, is the quadrupole formula \c{kludge, thorne}
\begin{equation}\label{quad}
\bar{h}^{ij}(t,\mathbf{x})=\frac{2}{D_L} \left[\ddot{I}^{ij}(t-D_L)\right] \quad \text{with} \quad I^{ij}=\int x^i x^j T^{tt}(t,\mathbf{x})\, d^3x.
\end{equation}
$D_L$ in the above equation is the EMRI's luminosity distance from the detector. The stress-energy tensor for the smaller compact object of mass $m$ is \c{thorne}
\begin{equation}
T^{t t}\left(t, x^i\right)=m \delta^3\left(x^i-Z^i(t)\right),
\end{equation}
where $Z^i(t)$ is the object's trajectory.This reduces Eq. (\ref{quad}) to the following expression:
\begin{equation}
h_{ij}=\frac{2m}{D_L}
\left(a_i x_j + a_j x_i + 2 v_i v_j \right),
\end{equation}
where $v_i$ and $a_i$ are, respectively, the velocity and acceleration of the orbiting object. We then project the above waveform onto the detector frame $(X,\,Y,\,Z)$ through the orthonormal basis
\begin{eqnarray}
\hat{e}_X &=& (\cos\zeta,-\sin\zeta,0),\\
\hat{e}_Y &=& (\sin\iota\sin\zeta,\cos\iota\cos\zeta,-\sin\iota),\\
\hat{e}_Z &=& (\sin\iota\sin\zeta,-\sin\iota\cos\zeta,\cos\iota),
\end{eqnarray}
where $\zeta$ is the pericenter's longitude and $\iota$ is the angle of inclination. Two polarization modes of GWs are \c{adia3, adia5}
\begin{eqnarray}
h_+ &=& \frac{1}{2}(e_X^i e_X^j - e_Y^i e_Y^j) h_{ij},\\
h_\times &=& \frac{1}{2}(e_X^i e_Y^j - e_Y^i e_X^j) h_{ij}.
\end{eqnarray}
To illustrate gravitational waveforms for different periodic orbits and to probe the impact of magnetic charge and DM on them, we consider an EMRI system where the masses of the smaller object and the central supermassive BH are $m=10M_\odot$ and $M=10^7 M_\odot$, respectively. We assume $\iota=\zeta=\pi/4$. Fig. (\ref{120}) illustrates the zoom-whirling motion of the smaller object and its correspondence with the emitted gravitational waveform. Using different colors for different segments is particularly helpful in highlighting how strong field influences GWs. Initially, the object travels in a highly elliptical zoom trajectory far from the BH, where the gravitational field is relatively weak; as a result, the corresponding segment of the waveform has lower amplitude and is less oscillatory. As the object nears perihelion close to the BH, it experiences strong acceleration and whirls around the BH. This whirl phase generates the central portion of the GW where the amplitude peaks and the oscillation is highest. Again, as the object returns to its initial position, the gravitational field decreases, and with it, the oscillatory nature of the GW decreases. \\
Fig. (\ref{413a}) demonstrates how the periodic orbit and gravitational waveforms vary with magnetic charge for the $(z,\,w,\,v)=(4,1,3)$ configuration. As we increase $g$, the periapsis shifts, albeit by a small amount, inward, and hence the gravitational field experienced by the smaller object increases, which is also reflected in the effective potential whose peak's height and well's depth increase with $g$. This results in a larger GW amplitude for larger values of $g$. There is also a phase shift which suggests a decrease in oscillation period with increasing $g$, implying shorter orbital timescale.\\
Fig. (\ref{413b}) graphically elucidates the effect of DM on the periodic orbit and gravitational waveforms for the $(z,\,w,\,v)=(4,1,3)$ configuration. Unlike in the case of magnetic charge, we observe a more pronounced impact of DM here. Increasing the DM parameter decreases the spatial extension of the zoom orbit, and the object whirls closer to the BH, thereby experiencing a stronger gravitational field while temporarily trapped near the BH. This is due to the strengthening of the effective potential as $\b$ increases. This enhances the oscillatory nature and amplitude of GWs as we increase $\b$. We observe that both the magnetic charge and DM amplify gravitational radiation, with the effect of DM more conspicuous.
\section{Conclusions}
The Bardeen BH is the first proposed regular BH and can also be obtained for a specific NED. This article credited Bardeen BH in the background of PFDM, arising from the minimal coupling between NED and matter. Modified field equations derived from the action were obtained and subsequently solved, yielding the desired static, spherically symmetric metric (\ref{final}). The resulting metric is no longer regular owing to the presence of DM. As expected, the limit $\b \rightarrow 0$ reduces the metric to that for the Bardeen BH, and a further limit $g \rightarrow 0$ yields the \s metric. The central theme of this article is to explore the combined effect of NED and DM on periodic orbits and gravitational waveforms, where an EMRI system with the central supermassive object being a BPFDM BH was considered. To this end, the effective potential governing the motion of test particles is the key that we obtained from the Lagrangian and the subsequent equations of motion. By imposing appropriate conditions on the effective potential, we obtained constants of motion and the positions of the MBO and ISCO. At ISCO, the effective potential exhibited a single extremum. As we increase the angular momentum, the potential exhibits two extrema - one maximum for the inner unstable orbit and one minimum for the outer stable orbit. The turning point was when the angular momentum reached $L_{MBO}$, beyond which no bound orbit could exist. The radii of both MBO and ISCO were found to be adversely affected by an increase in either $g$ or $\b$, implying strengthening of effective potential.\\

In EMRIs, the stellar object inspirals toward the central supermassive BH and loses energy via GW radiation, where periodic orbits serve as transition states. Periodic orbits are bound orbits where the particle's orbit closes upon itself after some time. Since the particle's motion is assumed to be confined to the equatorial plane, a rational number $q$ can be attributed to each periodic orbit. We observe that the onset of extreme whirling motion, where $q$ diverges, occurs at lower angular momentum or higher energy as we increase either the magnetic charge or the DM parameter. We also considered the $(z,w,v)$ taxonomy to systematically organize periodic orbits. For a fixed $(z,w,v)$ configuration, the energy and angular momentum required decrease with an increase in either $g$ or $\b$. Periodic orbits displayed in Fig. (\ref{tra}) reveal how varying configuration changes orbital complexity. While increasing $z$ increases the number of elliptical orbits the particle traces before returning to its initial position, increasing $w$ increases the number of whirls it takes around the central BH. Thus, the particle stays trapped for a longer period in the strong-field region near the BH as the whirl number $w$ increases. \\

Using the adiabatic approximation and the numerical "Kludge" method, we generated gravitational waveforms from EMRIs, in which the impact of zoom and whirl motions on different segments of the GW was graphically elucidated. When the smaller object moves in a zoom orbit, it travels in a weak field; consequently, the corresponding segment of the GW shows lower amplitude and a subdued oscillatory behavior. As the object approaches its periapsis near the BH, it gets temporarily trapped in a strong field and experiences high acceleration. Its corresponding segment displays high frequency and the amplitude peaks here. We also explored how variations in $g$ or $\b$ affect GW. For both these parameters, the amplitude increases and the orbital time period decreases with an increase in either of these parameters without any change in the fundamental waveform structure. The effect of DM, however, is significant, whereas the magnetic charge's impact is not that pronounced. These observations clearly demonstrate a distinct imprint of the interplay between NED and DM on orbital dynamics and emitted GWs. The efficacy of our model may be tested using future observations of GW emission from EMRIs.

\end{document}